\documentclass[bookmarks=true]{JHEP3}
\usepackage{amsmath}
\usepackage{amsfonts}
\usepackage{latexsym}
\usepackage[dvips]{graphicx}
\usepackage{bm}
\usepackage{slashed}
\usepackage{epsfig}
\newcommand{\starco}{\stackrel{\star}{,}}
\newcommand{\inv}[1]{\frac{1}{#1}}
\def\r{\rho}
\def\s{\sigma}
\def\t{\tau}
\def\e{\epsilon}
\renewcommand{\hepth}[1]{\href{http://www.arXiv.org/abs/hep-th/#1}{\tt hep-th/#1}}
\renewcommand{\grqc}[1]{\href{http://www.arXiv.org/abs/gr-qc/#1}{\tt gr-qc/#1}}
\title{Divergences in Non-Commutative Gauge Theories with the Slavnov Term}
\author{Daniel~N.~Blaschke\thanks{work supported by "Fonds zur F\"orderung der Wissenschaftlichen Forschung" (FWF) under contract P15463-N08}~, Stefan~Hohenegger\footnotemark[1]~\footnotemark[2]~, Manfred~Schweda\footnotemark[1]~\\
\footnotemark[1]~Institute for Theoretical Physics, Vienna University of Technology\\Wiedner Hauptstrasse 8-10, A-1040 Vienna, Austria\vspace{0.1cm}\\
\footnotemark[2]~Department of Physics, CERN -- Theory Division\\CH-1211 Geneva 23, Switzerland\vspace{0.3cm}\\
E-mail: \email{blaschke@hep.itp.tuwien.ac.at}, \email{stefan.hohenegger@cern.ch}, \email{mschweda@tph.tuwien.ac.at}}
\abstract{The divergence structure of non-commutative gauge field theories (NCGFT) with a Slavnov extension~\cite{Slavnov,slavnov2} is examined at one-loop level with main focus on the gauge boson self-energy. Using an interpolating gauge we show that even with this extension the quadratic IR divergence of the gauge boson self-energy is independent from a covariant gauge fixing as well as from an axial gauge.

The proposal of Slavnov is based on the fact that the photon propagator shows a new transversality condition with respect to the IR dangerous terms. This novel transversality is implemented with the help of a new dynamical multiplier field. However, one expects that in physical observables such contributions disappear. A further new feature is the existence of new UV divergences compatible with the gauge invariance (BRST symmetry). We then examine two explicit models with couplings to fermions and scalar fields.}
\preprint{\href{http://www.arXiv.org/abs/hep-th/0510100}{hep-th/0510100}\\CERN-PH-TH/2005-185}
\graphicspath{{figures/}}
\begin{document}
\section{Introduction}
The idea that some minimum length should exist, naturally arises as soon as one tries to combine both gravity and the quantum field theory of matter~\cite{Doplicher}: If the uncertainties $\Delta x_\mu$ in the measurement of coordinates of a point particle become sufficiently small, which is the case near the Planck length ($\lambda_p\simeq10^{-33}\text{cm}$), the gravitational field generated by the measurement will become so strong as to prevent light or other signals from leaving the region in question. To avoid black holes from being produced in the course of measurement one is tempted to introduce quantum, or non-commutative, space-time. Apart from this motivation, there is also the fact that ultraviolet divergent terms appear in quantum field theories due to point particles interacting locally. This also suggests that at very small distances physical laws must be modified in such a way that interactions become \emph{non-local}.

This can be achieved by introducing quantized non-commutative space-time~\cite{Snyder, connes} in order to describe quantum field models with a quantum structure of space-time pertubatively. In the last decade this new concept has received enormous interest since it has recently been discovered that non-commutativity between coordinates appears in open string theories with a B-field background as well as in toroidal compactification of Matrix Theory~\cite{connes2, Mora, Witten}.

Among the various possibilities for a quantized space-time the simplest way to formulate non-commutative quantum field theories (NCQFT) is to introduce a $\theta$ deformed space-time in the sense of Filk~\cite{Filk}, meaning that this NCQFT may be expressed in terms of ordinary commuting space-time coordinates instead of operator valued objects. This concept demands the field products of the action describing any NCQFT to be replaced by the so-called {\em Weyl-Moyal star products} (see Section~\ref{starprod} for a review). Unfortunately, field theories formulated in this way suffer from UV/IR mixing~\cite{Susskind}.

A first idea to get rid of these new divergences was to expand the star products in the action up to a given order (for simplicity usually first order) in $\theta$. In doing this one arrives at the so-called {\em Seiberg Witten map}~\cite{Witten}, a formulation in which gauge field theories are manifestly IR-{\em finite} in the sense of UV/IR mixing. Only the usual UV divergences are present. In this sense, a $\theta$-expanded deformed non-commutative Maxwell theory was discussed in~\cite{bichl-SWmap}. The discovery of Wulkenhaar et al.~\cite{wulkenh1,bichl-SWmap} that such theories are {\em non-renormalizable} if one also adds fermions to the pure gauge sector was surely a drawback.

However, Slavnov~\cite{Slavnov,slavnov2} suggested a different way of dealing with the arising IR singularities: Instead of expanding the theory in $\theta$ he simply added a further term in the action in order to modify the Feynman rules in an appropriate way. Our aim in this study is to discuss the effect of this 'Slavnov extension' through explicit one-loop calculations in non-commutative quantum electrodynamics coupled to fermions as well as to scalar fields.

The paper is organized as follows: After a brief review of the basic properties of $\theta$ deformed space-time we discuss the basic idea of Slavnov extended non-commutative $U(1)$ gauge theories in $d$ dimensions. In order to discuss gauge (in)dependence of the appearing highest order IR divergences we use an interpolating gauge~\cite{su-long,schweda, balasin, paper3}. In Section~\ref{qed3} we continue with coupling the model to fermions. In doing so we will go to $2+1$ dimensional Minkowski space since this number of dimensions is particularly interesting in comparison to the commutative model: Commutative QED in three dimensions is finite (in dimensional regularization) but as will be discussed, UV/IR mixing nevertheless leads to an IR divergence in the non-commutative model. Our aim, therefore, is to find out what happens when extending the model with Slavnov's additional term in the action.

Finally, Section~\ref{sqed} is going to deal with $4$-dimensional scalar QED (NCSQED). In order to be able to be more flexible in choosing the explicit form of $\theta$ deformed space-time without getting into trouble with causality\footnote{Causality is violated if time does not commute with space (i.e.~$\theta^{0i}\neq0$). In that case, the scattering matrix is no longer unitary~\cite{Seiberg, Connes}.}, Euclidean space will this time be discussed.
\section{$\theta$ deformed Space-Time}\label{starprod}
Following the work of Filk~\cite{Filk}, where the (commuting) coordinates of flat Minkowski space $\mathbb{M}^d$ are replaced by Hermitian operators $\hat{x}^\mu$ (with $\mu=0,1,\ldots,(d-1)$), we consider a canonical structure defined by the following algebra:
\begin{align}\label{comm}
\left[\hat{x}^\mu,\hat{x}^\nu\right]&=i\theta^{\mu\nu},\nonumber\\
\left[\theta^{\mu\nu},\hat{x}^\rho\right]&=0.
\end{align}
The matrix $\theta^{\mu\nu}$ is real, constant and antisymmetric. In natural units, where $\hbar=c=1$, its mass dimension $[\theta]=-2$. At this point one also has to mention that the commutation relations (\ref{comm}) between the coordinates explicitly break Lorentz invariance~\cite{lorentzinv1,lorentzinv2,lorentzinv3}. Furthermore, we call a space with the above commutation relations a {\em non-commutative} space $\mathbb{M}^d_{NC}$.

In order to construct the perturbative field theory formulation, it is more convenient to use fields $\Phi(x)$ (which are functions of ordinary commuting coordinates) instead of operator valued objects like $\hat{\Phi}(\hat{x})$. To be able to pass to such fields, in respecting the properties (\ref{comm}), one must redefine the multiplication law of functional (field) space. This new multiplication is induced by the algebra (\ref{comm}) through the so-called Weyl-Moyal correspondence~\cite{Bichl2, Minwalla}. For a scalar field one has the following definitions:
\begin{align}
\hat{\Phi}(\hat{x})&\longleftrightarrow\Phi(x),\nonumber\\
\hat{\Phi}(\hat{x})&=\int\frac{d^dk}{(2\pi)^d}e^{ik\hat{x}}\widetilde{\Phi}(k),\nonumber\\
\widetilde{\Phi}(k)&=\int d^dxe^{-ikx}\Phi(x),
\end{align}
where $k$ and $x$ are real variables. For two arbitrary scalar fields $\hat{\Phi}_1$, $\hat{\Phi}_2$ one therefore has\footnote{One has to use the Baker-Campbell-Hausdorff formula, as well as relation (\ref{comm}).}
\begin{align}
\hat{\Phi}_1(\hat{x})\hat{\Phi}_2(\hat{x})&=\int\frac{d^dk_1}{(2\pi)^d}\int\frac{d^dk_2}{(2\pi)^d}\widetilde{\Phi}_1(k_1)\widetilde{\Phi}_2(k_2)e^{ik_1\hat{x}}e^{ik_2\hat{x}}\nonumber\\
&=\int\frac{d^dk_1}{(2\pi)^d}\int\frac{d^dk_2}{(2\pi)^d}\widetilde{\Phi}_1(k_1)\widetilde{\Phi}_2(k_2)e^{i(k_1+k_2)\hat{x}-\frac{1}{2}\left[\hat{x}^\mu,\hat{x}^\nu\right]k_{1,\mu}k_{2,\nu}}.\label{staroperatfull}
\end{align}
Hence one has the following correspondence:
\begin{align}
\hat{\Phi}_1(\hat{x})\hat{\Phi}_2(\hat{x})\longleftrightarrow\Phi_1(x)\star\Phi_2(x),
\end{align}
with
\begin{align}\label{star2}
\Phi_1(x)\star\Phi_2(x)=e^{\frac{i}{2}\theta^{\mu\nu}\partial^x_\mu\partial^y_\nu}\Phi_1(x)\Phi_2(y)\big|_{x=y},
\end{align}
where relation (\ref{comm}) was used to replace the commutator in the exponent of (\ref{staroperatfull}). This means that we can work on a usual commutative space for which the multiplication operation is modified by the so-called {\em star product} (\ref{star2}). For the ordinary commuting coordinates this implies\footnote{The Weyl bracket is defined as $[A\starco B]=A\star B-B\star A$.}
\begin{align}\label{comm2}
\left[x^\mu\starco x^\nu\right]&=i\theta^{\mu\nu},\nonumber\\
\left[\theta^{\mu\nu}\starco x^\rho\right]&=0.
\end{align}
A natural generalization of (\ref{star2}) is given by
\begin{align}\label{defstar}
\Phi_1(x)\star\Phi_2(x)\star\cdots\star\Phi_m(x)=&\,\int\frac{d^dk_1}{(2\pi)^d}\int\frac{d^dk_2}{(2\pi)^d}\cdots\int\frac{d^dk_m}{(2\pi)^d}e^{i\sum\limits_{i=1}^mk_i^\mu x_\mu}\nonumber\\
&\,\times\widetilde{\Phi}_1(k_1)\widetilde{\Phi}_2(k_2)\cdots\widetilde{\Phi}_m(k_m)e^{-\frac{i}{2}\sum\limits_{i<j}^mk_i\times k_j},
\end{align}
where $k\times k'$ is an abbreviation for $k\times k'\equiv k_\mu\theta^{\mu\nu}k'_\nu\equiv k_{\mu}\tilde{k}'^{\mu}$. Furthermore, one can easily verify the following properties of the star product
\begin{subequations}
\begin{align}\label{starprop1}
\int d^dx\left(\Phi_1\star\Phi_2\right)(x)&=\int d^dx\ \Phi_1(x)\Phi_2(x),\\\label{starprop2}
\int d^dx\left(\Phi_1\star\Phi_2\star\cdots\star\Phi_m\right)(x)&=\int d^dx\left(\Phi_2\star\cdots\star\Phi_m\star\Phi_1\right)(x),\\\label{starprop3}
\frac{\delta}{\delta\Phi_1(y)}\int d^dx\left(\Phi_1\star\Phi_2\star\cdots\star\Phi_m\right)(x)&=\left(\Phi_2\star\cdots\star\Phi_m\right)(y).
\end{align}
\end{subequations}
Equations (\ref{defstar}) and (\ref{starprop1}) demonstrate that in a $\theta$ deformed quantum field theory the free field part is not modified and therefore the corresponding propagators remain unchanged. Only the interaction terms in the action are equipped with additional phases leading to complete new features in the perturbative realization of the corresponding non-commutative quantum field theories (NCQFT) now containing planar and non-planar graphs.
\section{Non-Commutative Gauge Theories in $d$ Dimensions}\label{noncomm-MT}
\subsection{Basics}
Before we consider interactions with matter, we first want to clarify some details in the pure gauge field sector: Starting from an ordinary gauge theory based on a Lie-group with a matrix-valued gauge field $A_\mu(x)=A_\mu^aT^a$, where $T^a$ are generators of some gauge group, the na\"ive correspondence principle yields at the classical level
\begin{align}
\Gamma^{(0)}_{\text{gauge}}=-\frac{1}{4}\int d^dxF_{\mu\nu}\star F^{\mu\nu},
\end{align}
with
\begin{align}
F_{\mu\nu}=\partial_{\mu}A_{\nu}-\partial_{\nu}A_{\mu}-ig[A_{\mu}\starco A_{\nu}]\,,
\end{align}
for the pure gauge sector. The infinitesimal gauge transformation is given by
\begin{align}\label{gaugetrafo}
\delta_\Lambda A_\mu=D_\mu\Lambda=\partial_\mu\Lambda-ig[A_{\mu}\starco \Lambda],
\end{align}
with $\Lambda=\Lambda^aT^a$ some $x$-dependent gauge parameter. One has to stress that taking the gauge fields valued on an arbitrary Lie-algebra is not in general consistent with the non-commutative deformation. In order to see this we expand equation (\ref{gaugetrafo}):
\begin{align}\label{gaugetrafo2}
\delta_\Lambda A_\mu=T^a\partial_\mu\Lambda^a-\frac{ig}{2}[T^a,T^b]\{A_\mu^a\starco\Lambda^b\}-\frac{ig}{2}\{T^a,T^b\}[A_\mu^a\starco\Lambda^b].
\end{align}
Obviously, the gauge transformation not only depends on the commutator of the generators but also the {\em anticommutator} $\{T^a,T^b\}$ due to the non-commutative character of the star product. This means that $\{T^a,T^b\}$ must be a linear combination of $T^c$ so that the gauge transformation is closed. Unfortunately, in a non-commutative theory this is not the case for $SU(N)$, but $\{T^a,T^b\}$ is an element of the Lie-algebra in the case of the $U(N)$ group. Furthermore, equation (\ref{gaugetrafo2}) shows that even the $U(1)$ gauge theory has non-Abelian character on $\theta$ deformed space-time entailing a BRST quantization procedure with ghost field $c$ replacing the infinitesimal gauge parameter $\Lambda$ and by introducing the anti-ghost field $\bar{c}$.
\subsubsection{The Classical Action}
Thus, in considering a non-commutative $U(1)$ Maxwell theory at the classical level, we have the following action in an $d$-dimensional space-time including the term proposed by A.A. Slavnov~\cite{Slavnov,slavnov2}
\begin{align}\label{LagrangianMT}
\Gamma^{(0)}_{\text{MT}}=\int d^dx\left(-\frac{1}{4}F_{\mu\nu}\star F^{\mu\nu}+\frac{\alpha}{2}B\star B+B\star N^{\mu}A_{\mu}-\bar{c}\star N^{\mu}D_{\mu}c+\frac{\lambda}{2}\star\theta^{\mu\nu}F_{\mu\nu}\right)(x),
\end{align}
with\footnote{We do not consider light like gauges $n^2=0$ in this study.}
\begin{align}
D_{\mu}c=\partial_{\mu}c-ig[A_{\mu}\starco c],\nonumber\\
N_\mu=\partial_\mu-\xi\frac{(n\partial)}{n^2}n_\mu.
\end{align}
$N_\mu$ describes a gauge fixing which interpolates between a covariant gauge and an axial gauge. $\xi$ is a real gauge parameter taking values between $(-\infty,+1)$ and $n_\mu$ is a constant gauge vector. Such an interpolating gauge was originally proposed in~\cite{su-long} and used in~\cite{schweda, balasin} as well as in a recent paper~\cite{paper3}. The last term in (\ref{LagrangianMT}) describes the 'Slavnov extension' as stated in the Introduction of this paper and which will therefore be called the {\em Slavnov term}. It contains the new multiplier field $\lambda(x)$. The model therefore contains two multiplier fields $B$ and $\lambda$ entailing the following constraint equations:
\begin{align}\label{constraint1}
\frac{\delta\Gamma^{(0)}_{\text{MT}}}{\delta B}&=\alpha B+N^\mu A_\mu=0,\\\label{constraint2}
\frac{\delta\Gamma^{(0)}_{\text{MT}}}{\delta \lambda}&=\frac{1}{2}\theta^{\mu\nu}F_{\mu\nu}=0.
\end{align}
Equation (\ref{constraint1}) is devoted to fix the gauge with the two gauge parameters $\alpha$ and $\xi$ and a gauge vector $n_\mu$. Some choices are quite prominent in the literature:
\begin{itemize}
\item $\xi=0$ and $\alpha$=0, normally called Landau gauge
\item $\xi=0$ and $\alpha$=1, usually known as Feynman gauge
\item $\xi\to-\infty$ (or $N_\mu=n_\mu$) and $\alpha=0$, leading to the homogeneous axial gauge. 
\end{itemize}
In this paper, however, we will use generic gauge parameters in order to investigate the dependence of the highest IR poles of $\xi$ and $\alpha$.

Relation (\ref{constraint2}), on the other hand, induces a first class constraint (i.e.~\cite{Dirac}) and changes the gauge field propagator in a drastic manner: The propagator becomes transverse with respect to $\tilde{k}^\mu$. In momentum space this means:
\begin{align}\label{photon-prop}
i\Delta^{A}_{\mu\nu}(k)&=-\frac{i}{k^2}\left[g_{\mu\nu}-ak_\mu k_\nu+b(n_\mu k_\nu+k_\mu n_\nu)-b'\left(\tilde{k}_\mu k_\nu+k_\mu\tilde{k}_\nu\right)-\frac{\tilde{k}_\mu\tilde{k}_\nu}{\tilde{k}^2}\right]
\end{align}
with
\begin{align}\label{defabb'}
a&=\frac{(1-\alpha)k^2-\zeta^2(nk)^2\left[n^2-\frac{(n\tilde{k})^2}{\tilde{k}^2}\right]}{\left[k^2-\zeta(nk)^2\right]^2},\nonumber\\
b&=\frac{\zeta(nk)}{k^2-\zeta(nk)^2}\quad,\quad b'=\frac{(n\tilde{k})}{\tilde{k}^2}b,\nonumber\\
\zeta&=\frac{\xi}{n^2}\quad,\quad n^2\neq0
\end{align}
In the limit $\zeta \rightarrow 0$ $(\xi \rightarrow 0)$ one recovers the gauge field propagator for a covariant gauge fixing characterized by the gauge parameter $\alpha$:
\begin{align}
i\Delta^{cov}_{\mu \nu} = -\frac{i}{k^2} \left[g_{\mu \nu} - (1 - \alpha) 
\frac{k_\mu k_\nu}{k^2} -\frac{\tilde{k}_\mu\tilde{k}_\nu}{\tilde{k}^2}\right].
\end{align}
In the limit $\zeta \rightarrow - \infty$ ($\xi \rightarrow - \infty$) and $n^2 \ne 0$ one has the corresponding gauge field propagator in the axial gauge:
\begin{align}
i\Delta^{ax}_{\mu \nu} =  \frac{-i}{k^2} \left[g_{\mu \nu}+
\left(n^2-\frac{(n\tilde{k})^2}{\tilde{k}^2}\right) \frac{k_\mu k_\nu}{(nk)^2} -
\frac{n_\mu k_\nu + n_\nu k_\mu}{(nk)} +\frac{(n\tilde{k})}{\tilde{k}^2}\frac{\left(\tilde{k}_\mu k_\nu+k_\mu\tilde{k}_\nu\right)}{(nk)}-\frac{\tilde{k}_\mu\tilde{k}_\nu}{\tilde{k}^2}\right].
\end{align}
From (\ref{photon-prop}) follows
\begin{align}\label{sl-trick}
\tilde{k}^{\mu}\Delta^{A}_{\mu\nu}(k)&=-\inv{k^2}\left[\tilde{k}_\nu+b(\tilde{k}n)k_\nu-b'\tilde{k}^2k_\nu-\tilde{k}_\nu\right]=0,
\end{align}
where the definition of $b'$ in (\ref{defabb'}) was used. This new kind of transversality is actually encoded in equation (\ref{constraint2}) if one considers only the bi-linear parts of the action responsible for calculating the gauge field propagator and will be very useful in avoiding the UV/IR mixing at higher loop order: It is already known from earlier studies (e.g.~\cite{hayakawa}) that the dangerous IR singularities of self-energy insertions of the gauge boson are of the form\footnote{It has in fact been shown in the literature that these IR divergences are gauge independent~\cite{ruiz, paper3}.}
\begin{equation}\label{IRdiv}
\Pi^{\text{IR}}_{\mu\nu}(k)\propto \frac{\tilde{k}_{\mu}\tilde{k}_{\nu}}{(\tilde{k}^2)^{\frac{d}{2}}},
\end{equation} 
where $d$ denotes the number of space-time dimensions. If one inserts these structures into higher order diagrams the loop integrations lead to problems around $k=0$ if $n>2$. With the new transversality property of the propagator this problem is circumvented as is shown with the help of Figure~\ref{fig:insertion}.
\FIGURE[h]{\parbox[h]{15cm}{\centering\epsfig{file=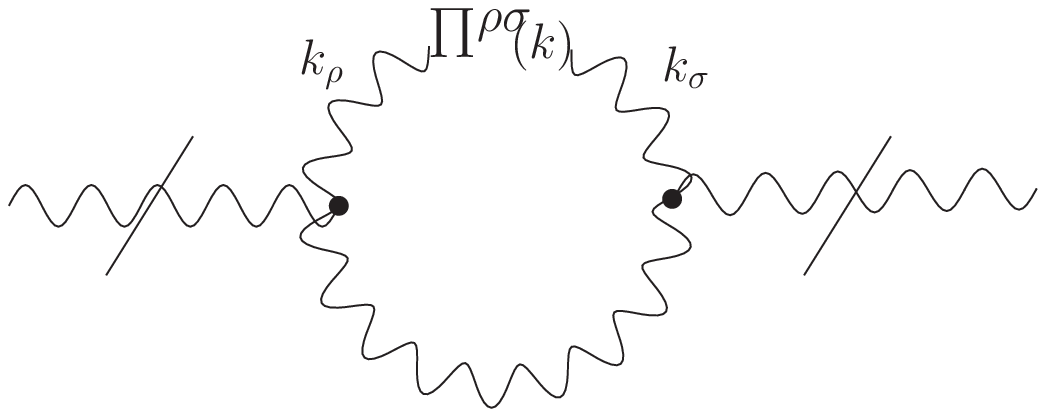, width=7cm}
\caption{A possible example for an insertion for (amputated) higher order graphs}\label{fig:insertion}}}
As one clearly sees, the inserted self-energy is multiplied by two internal propagators in the way $\Delta^{A}_{\mu\rho}(k)\Pi^{\rho\sigma}(k)\Delta^{A}_{\sigma\nu}(k)$ before integrating. Especially, for the IR divergent parts we have
\begin{equation}
\Delta^{A}_{\mu\rho}(k)\frac{\tilde{k}^{\rho}\tilde{k}^{\sigma}}{(\tilde{k}^2)^{\frac{d}{2}}}\Delta^{A}_{\sigma\nu}(k)=0.
\end{equation}
This means the insertions of the IR dangerous parts vanish even without integration. This procedure, of course, requires that no further IR divergences with a different tensor structure appear in the model. There are some convincing arguments in \cite{Slavnov} that only terms like (\ref{IRdiv}) are obtained, though an explicit one-loop calculation is missing. It is one of the tasks of this paper to show the outcome of such calculations.

There is another new feature in connection with the multiplier field $\lambda$: It becomes a dynamical field with a non-vanishing propagator and also induces new polynomial interactions with the gauge fields which might cause further problems which will also be discussed in this paper.
\subsubsection{The BRST Transformations}
In order to quantize the model consistently one has to replace the infinitesimal gauge transformation with an appropriate BRST transformation. Seeking adequate transformations rendering (\ref{LagrangianMT}) invariant we can in principle use similar rules as in ordinary commutative, non-Abelian field theories except for replacing ordinary products by star products, namely
\begin{subequations}\label{BRS}
\begin{align}
& sA_\mu = D_\mu c= \partial_\mu c - ig [A_\mu \starco c], && sc=igc \star c, \nonumber\\
& s \bar c = B, && sB = 0, \nonumber \\
&s^2\phi=0, \hspace{1cm} \text{for}\ \phi=\{A^{\mu},B,c,\bar{c}\}. 
\label{brs1}
\end{align}
Also for the new field $\lambda(x)$ we have to find a corresponding BRST transformation. But since the field tensor $F_{\mu\nu}$ transforms covariantly ($sF_{\mu\nu}=-ig[F_{\mu\nu}\starco c]$), as can easily be checked, it is not difficult to see that the relation
\begin{equation}\label{BRSL}
s\lambda=-ig[\lambda\starco c],
\end{equation}
\end{subequations}
renders the Slavnov term invariant. It is also rather easy to check that with these relations the BRST operator is nilpotent.
\subsubsection{The Slavnov-Taylor Identities}
In order to describe the symmetry content of the non-linear and supersymmetric BRST-symmetry by a corresponding Ward-identity --- the so-called Slavnov-identity~\cite{piguet,boresch} --- one has to introduce BRST-invariant unquantized external sources for the non-linear parts of the equations (\ref{BRS}). This leads to the introduction of 
\begin{align}
\Gamma_{\text{ex}}=\int d^dx\left(\rho^{\mu}\star sA_{\mu}+\sigma\star sc+\gamma\star s\lambda\right)(x),
\end{align}
with $s\rho^{\mu}=s\sigma=s\gamma=0$. The total new action therefore becomes
\begin{align}
\Gamma^{(0)}&=\Gamma^{(0)}_{\text{MT}}+\Gamma_{\text{ex}}=\nonumber\\
&=\int d^dx\bigg(-\frac{1}{4}F_{\mu\nu}\star F^{\mu\nu}+\frac{\alpha}{2}B\star B+B\star N^{\mu}A_{\mu}-\bar{c}\star N^{\mu}D_{\mu}c+\frac{\lambda}{2}\star\theta^{\mu\nu}F_{\mu\nu}+\nonumber\\
&\hspace{2cm}+\rho^{\mu}\star sA_{\mu}+\sigma\star sc+\gamma\star s\lambda\bigg)(x).
\end{align}
Hence, at the classical level, one has the following non-linear identity for the classical vertex functional
\begin{align}\label{Taylor}
\mathcal{S}\left(\Gamma^{(0)}\right)=\int d^dx\left(\frac{\delta\Gamma^{(0)}}{\delta\rho^{\mu}}\star\frac{\delta\Gamma^{(0)}}{\delta A_{\mu}}+\frac{\delta\Gamma^{(0)}}{\delta\sigma}\star\frac{\delta\Gamma^{(0)}}{\delta c}+\frac{\delta\Gamma^{(0)}}{\delta\gamma}\star\frac{\delta\Gamma^{(0)}}{\delta \lambda}+B\star\frac{\delta\Gamma^{(0)}}{\delta \bar{c}}\right)=0.
\end{align}
This equation describes the symmetry content with respect to (\ref{BRS}). Together with the linearized BRST operator \cite{piguet,boresch}
\begin{align}
\mathcal{S}_{\Gamma^{(0)}}=\int d^dx&\bigg(\frac{\delta\Gamma^{(0)}}{\delta\rho^{\mu}}\star\frac{\delta}{\delta A_{\mu}}+\frac{\delta\Gamma^{(0)}}{\delta A_{\mu}}\star\frac{\delta}{\delta \rho^{\mu}}+\frac{\delta\Gamma^{(0)}}{\delta\sigma}\star\frac{\delta}{\delta c}+\frac{\delta\Gamma^{(0)}}{\delta c}\star\frac{\delta}{\delta \sigma}+\nonumber\\
&+\frac{\delta\Gamma^{(0)}}{\delta\gamma}\star\frac{\delta}{\delta \lambda}+\frac{\delta\Gamma^{(0)}}{\delta\lambda}\star\frac{\delta}{\delta \gamma}+B\star\frac{\delta}{\delta \bar{c}}\bigg),
\end{align}
one gets from (\ref{Taylor})
\begin{align}\label{Taylorid}
\frac{\delta}{\delta A_{\rho}(y)}\mathcal{S}(\Gamma^{(0)})=\mathcal{S}_{\Gamma^{(0)}}\frac{\delta\Gamma^{(0)}}{\delta A_{\rho}(y)}=0.
\end{align}
When taking the functional derivative of (\ref{Taylorid}) with respect to $c$ and then setting all fields to zero one obtains the transversality condition for the one-particle irreducible (1PI) two-point graph
\begin{align}\label{transversality}
\partial_{\mu}^{y}\frac{\delta^2\Gamma^{(0)}}{\delta A_{\mu}(y)\delta A_{\nu}(y)}=0.
\end{align}
The central task of the perturbative analysis is to study the behaviour of the symmetry content in the presence of radiative corrections. One important question is the validity of (\ref{transversality}) at the perturbative level.
\subsubsection{The free theory}\label{free-theory}
A further task to be done is to consider the free theory, especially the behaviour of the newly introduced field $\lambda$. The free field equations derived from the action (\ref{LagrangianMT}) are given by
\begin{subequations}
\begin{align}\label{free-1a}
\partial^\mu (\partial_\mu A_{\nu}-\partial_\nu A_\mu)-N_\nu B+\widetilde{\partial}_\nu\lambda&=0,\\\label{free-1b}
\widetilde{\partial}^\mu A_\mu&=0,\\\label{free-1c}
N^\mu A_\mu+\alpha B&=0.
\end{align}
\end{subequations}
However, the solution of this system of differential equations is not so trivial and will be thoroughly discussed in our work in progress~\cite{workinprogress}.
\subsection{Gauge boson self-energy at the one-loop level}\label{boson-interpolating}
Compared to the non-commutative model without the Slavnov term, we have many additional contributions to the gauge boson self-energy (see Figure~\ref{fig:bosgraphs}) at one-loop level. (For the definition of the Feynman rules see Appendix \ref{app-feyn}.) 
\FIGURE[h]{\includegraphics[scale=0.6]{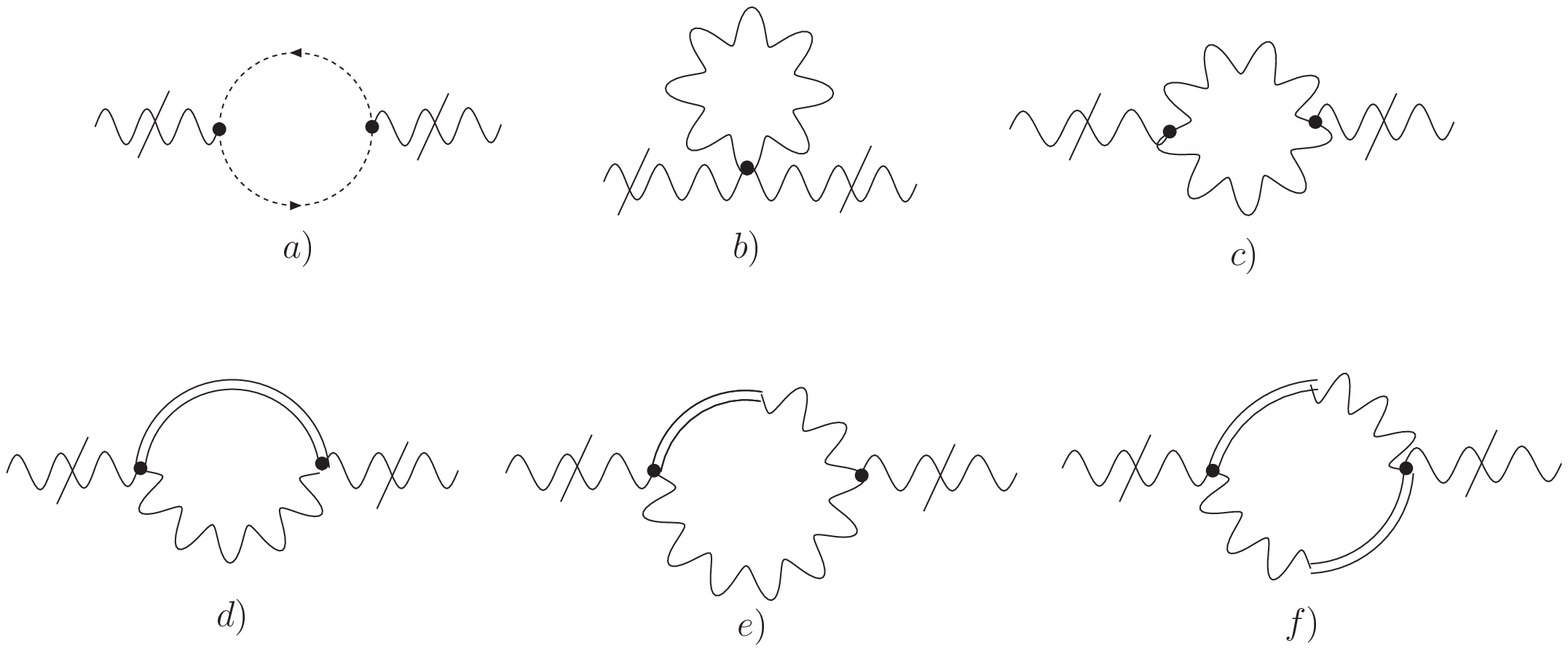}
\caption{gauge boson self-energy --- amputated graphs}
\label{fig:bosgraphs}}
In order to isolate the expected IR singularities of the non-planar sector we consider the following expansion of the momentum representation of the two point self-energy corrections in $d$-dimensions
\begin{align}
i\Pi_{\mu\nu} (p) =& \int d^dk I_{\mu \nu} (k, 0) \sin^2 \left(\frac{k\tilde{p}}{2}\right) + 
p^\rho \int d^dk \left(\frac{\partial}{\partial p^\rho} I_{\mu \nu} (k ,p)\Big|_{p=0}\right)
 \sin^2 \left(\frac{k\tilde{p}}{2}\right) + \nonumber\\
& + \inv 2 p^\sigma p^\rho \int d^dk 
\left(\frac{\partial^2}{\partial p^\sigma \partial p^\rho} I_{\mu \nu} (k ,p)\Big|_{p=0}\right)
 \sin^2 \left(\frac{k\tilde{p}}{2}\right)+\ldots\,, 
 \label{expansion}
\end{align}
where the worst expected IR divergence will appear in the non-planar part of the first term of this expansion. For the first term of equation (\ref{expansion}) one gets the following expression (see Appendix~\ref{app-graphs} for details):
\begin{align}\label{bosgraphs}
i\Pi^{\mu\nu}_{\text{IR}}(p)&\equiv\int d^dk I^{\mu \nu} (k, 0) \sin^2 \left(\frac{k\tilde{p}}{2}\right)=\nonumber\\
&=4g^2\int\frac{d^dk}{(2\pi)^d}\left\{(d-3)\left(2\frac{k^\mu k^\nu}{k^4}-\frac{g^{\mu\nu}}{k^2}\right)+\theta^{\mu\t}\left(\frac{g_{\t\s}}{\tilde{k}^2}-2\frac{\tilde{k}_\t\tilde{k}_\s}{\tilde{k}^4}\right)\theta^{\s\nu}\right\}\sin^2\left(\frac{k\tilde{p}}{2}\right),
\end{align}
where $d=g^\mu_{\ \mu}$ denotes the dimension of space-time. This result shows some very interesting features: First of all, we notice that all gauge-dependent terms have obviously cancelled leaving (\ref{bosgraphs}) completely {\em gauge independent}.

Furthermore, we notice that the first term vanishes in a three dimensional space-time. Let us compare (\ref{bosgraphs}) with the corresponding expression of a model {\em without} the Slavnov term: In this case one only has three graphs at the one-loop level, namely Figures~\ref{fig:bosgraphs}a),~\ref{fig:bosgraphs}b) and \ref{fig:bosgraphs}c). The result is (where $p$ describes the external momentum)
\begin{align}\label{bos-nosl}
i\Pi^{\mu\nu}_{\text{IR}}(p)&=4g^2\int\frac{d^dk}{(2\pi)^d}(d-2)\left[2\frac{k^\mu k^\nu}{k^4}-\frac{g^{\mu \nu}}{k^2}\right]\sin^2\left(\frac{k\tilde p}{2}\right),
\end{align}
showing that a 2-dimensional model (without the Slavnov Term) is IR {\em finite}. (\ref{bos-nosl}) is, of course, also gauge independent as has already been derived in the literature~\cite{ruiz,paper3,hayakawa}. Comparing (\ref{bosgraphs}) with (\ref{bos-nosl}) we see, that adding the Slavnov term changes the overall factor of the terms appearing in both models and, coming from the graphs depicted in Figures~\ref{fig:bosgraphs}d) and \ref{fig:bosgraphs}f), additional terms including $\theta_{\mu\nu}$ occur. (All other $\theta$-dependent terms, including those coming from the other graphs, cancel.)

An important question now is, whether these additional terms in the integrand also lead to IR divergent terms of the form (\ref{IRdiv}), since Slavnov's trick is based on the fact that the gauge propagator is transverse with respect to $\tilde{k}_\mu$ (see equation (\ref{sl-trick})). Actually, one sees immediately, that in case $\theta_{\mu\nu}$ does not have full rank, part of the integrand will be independent of certain directions and hence produce additional UV divergences. However, one may still hope, that these divergences are proportional to $\tilde{p}_\mu\tilde{p}_\nu$ in which case Slavnov's trick would still work. We will further discuss this problem in connection with two explicit models in Sections~\ref{qed3} and \ref{sqed}.
%
\section{Non-Commutative QED$_3$ Including Fermions}\label{qed3}
\subsection{The Model}
In this section we discuss non-commutative quantum electrodynamics in $2+1$ space-time dimensions coupled to fermions and compare the results with the commutative (and {\em finite}) counterpart~\cite{brazil}. In the previous section we have already shown that the worst divergences are gauge independent and hence we may choose a more convenient gauge for our explicit calculations. Choosing a covariant gauge the complete action of this model (including the Slavnov term) at the classical level reads
\begin{align}\label{S1}
\Gamma^{(0)}_{\text{QED}}=\int d^3x\bigg(&\bar{\psi}\star\left(i\slashed{\mathcal{D}}-m\right)\star\psi-\frac{1}{4}F^{\mu\nu}\star F_{\mu\nu}+B\star\partial^\mu A_\mu+\frac{\alpha}{2}B\star B-\nonumber\\&-\bar{c}\star\partial^\mu D_\mu c+\frac{\lambda}{2}\star\theta^{\mu\nu}F_{\mu\nu}\bigg)(x),
\end{align}
where again $B$ and $\lambda$ are Lagrange multiplier fields, $c$ is the ghost and $\bar{c}$ is the anti-ghost field. Because of (\ref{starprop1}) the star product may be dropped in the bilinear terms. The covariant derivative acting on a spinor is defined by
\begin{align}\label{covder}
\slashed{\mathcal{D}}=\gamma^\mu \mathcal{D}_\mu\ , \qquad \mathcal{D}_\mu=\partial_\mu-ig'A_\mu.
\end{align}
The set of $\gamma$ matrices has been chosen as
\begin{equation}
\gamma^\mu=\left(\sigma_1, i\sigma_2, -i\sigma_3\right),
\end{equation}
respecting the usual Clifford algebra
\begin{equation}\label{gammarel1}
\{\gamma^\mu,\gamma^\nu\}=2g^{\mu\nu},
\end{equation}
where $g^{\mu\nu}=\text{diag}\left(1,-1-1\right)$ is the 3-dimensional Minkowski metric. The $\sigma_i$ denote the well-known Pauli matrices\footnote{$\sigma_1=\left(\begin{array}{cc} 0 & 1\\1 & 0\end{array}\right), \quad \sigma_2=\left(\begin{array}{cc} 0 & -i\\i & 0\end{array}\right), \quad \sigma_3=\left(\begin{array}{cc} 1 & 0\\0 & -1\end{array}\right)$} and for the matrix describing non-commutativity $\theta^{\mu\nu}$ in this case we use
\begin{align}\label{theta}
\theta^{\mu\nu}=\theta\left(\begin{array}{ccc}
0 & 0 & 0\\
0 & 0 & 1\\
0 & -1 & 0\end{array}\right),
\end{align}
where $\theta$ is a real constant. In order to preserve causality, time still commutes with space (see (\ref{theta})). Furthermore, the action (\ref{S1}) is invariant under the BRST-transformations given in (\ref{BRS}) as well as
\begin{align}
&s\psi=ig'c\star\psi,\nonumber\\
&s\bar{\psi}=ig'\bar{\psi}\star c,
\end{align}
for the fermions, provided that $g'=g$, as can be easily checked using the Jacobi identity and the invariance of the (integrated) star product under cyclic permutations (\ref{starprop2}). Note, that in contrast to electroweak theory where one has no restriction for the value of $g'$, BRS invariance in non-commutative QED leads to the new limition $g'=g$ (see~\cite{hayakawa}).

From the Feynman rules displayed in Tables~\ref{tab:propMT} and~\ref{tab:vertexMT} in the Appendix one can derive the UV power counting formula in the usual way
\begin{align}\label{dgammasl1}
d(\gamma)=3-E_\psi-\frac{1}{2}E_A-\frac{1}{2}E_c-\frac{3}{2}E_{\lambda}-\frac{1}{2}E_g,
\end{align}
where $E_\psi$, $E_A$, $E_c$ and $E_{\lambda}$ denote the number of external fermion-, photon-, ghost- and $\lambda$-legs, respectively. Finally, $E_g$ denotes the number of coupling constants\footnote{remember $g'=g$} $g$ and $g'$ which are hence treated as external fields. Counting the coupling constants in a Feynman graph is equivalent to counting vertices. (Notice that in $2+1$ dimensions the coupling constant has mass dimension $[g]=1/2$.)

Obviously, $E_g$ gets larger with growing loop order (which is equivalent to an increasing number of vertices). Therefore, there is only a limited number of superficially divergent graphs in this model: Since every Feynman graph has $E_g\geq2$ ($E_g=2$ at one-loop order), the fermion self-energy ($E_\psi=2$) appears to be, at most, logarithmically divergent, whereas superficially, the gauge field ($E_A=2$) and ghost ($E_c=2$) self-energies appear to be linearly divergent. Further superficially (logarithmically) divergent graphs are those including three external photon lines ($E_A=3$, $E_g=3$) and the (lowest order) correction to the ghost-photon vertex ($E_c=2$, $E_A=1$, $E_g=3$). We will, however, concentrate on the self-energy graphs in this paper.
\subsection{One-Loop Calculations}
\subsubsection{The Fermion Self-Energy}
Now that we have all the building blocks necessary for doing loop calculations, we will first take a look at the fermion self-energy at one-loop level: 
\FIGURE[h]{\parbox{4.2cm}{\centering\includegraphics[scale=0.6]{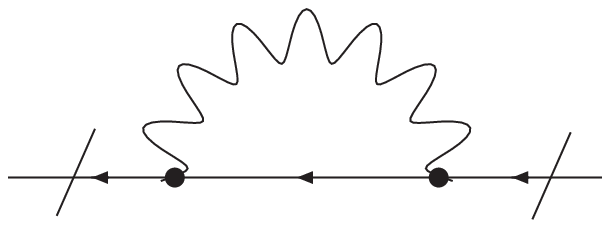}}
\caption{fermion self-energy --- amputated graph}
\label{fig:fermself}}
The only one-loop expression is represented in Figure~\ref{fig:fermself} and turns out to be completely independent of phases since phase factors at the two fermion vertices (see Table~\ref{tab:vertexMT}) cancel each other~\cite{hayakawa}. Hence, there is graphically {\em no modification} due to non-commutativity except that an additional piece in the gauge field propagator occurs due to the Slavnov term. In dimensional regularization the fermion self-energy is even {\em finite} and without the Slavnov extension one would get precisely the same expression as in the commutative model~\cite{brazil}.
\subsubsection{The Photon Self-Energy}
%
\FIGURE[h]{\includegraphics[scale=0.6]{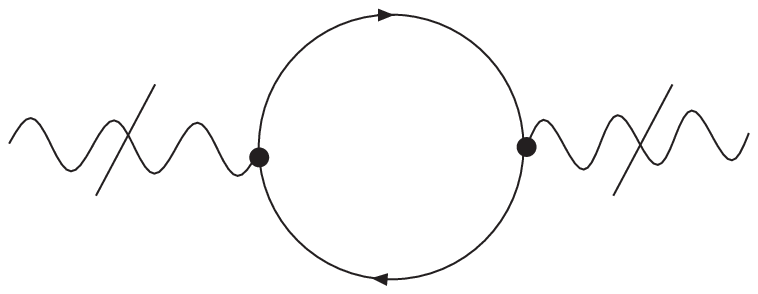}
\caption{amputated fermion loop graph}
\label{fig:fermloop}}
The photon self-energy on the other hand gets many additional contributions compared to the commutative model: As shown in Figure~\ref{fig:bosgraphs} and Figure~\ref{fig:fermloop} there are seven different Feynman graphs contributing to the vacuum polarization. The fermion loop displayed in Figure~\ref{fig:fermloop} is the only one that also appears in commutative QED and in fact this graph remains {\em unmodified}: As in the fermion self-energy the phases at the two vertices cancel each other~\cite{hayakawa} and the result, as known from the commutative model, is {\em finite}.

Since the Slavnov term leads to a change in the divergence structure of the remaining contributions we will first take a look at what happens without the Slavnov extension: In this case one only gets three additional graphs (Figures~\ref{fig:bosgraphs}a)-\ref{fig:bosgraphs}c)) which all contain phases of the form (see (\ref{bos-nosl}))
\begin{align}\label{phase}
\sin^2\left(\frac{k\times p}{2}\right)=\frac{1}{2}\left(1-\cos\left(k\times p\right)\right)=\frac{1}{4}\left(2-\sum_{\eta=\pm1}e^{i\eta k\tilde{p}}\right),
\end{align}
where $k$ denotes an internal momentum while $p$ stands for an external one. Hence, these graphs can be decomposed into finite planar parts which are independent of phases and non-planar parts which turn out to be linearly infrared divergent. So in contrast to the commutative model non-commutative $2+1$ dimensional QED is {\em not} finite but suffers from UV/IR mixing. The sum of these IR divergent terms takes the (transversal) form at the one-loop level
\begin{eqnarray}\label{IR1}
\Pi^{\mu\nu}_{\text{IR-divergent}}(p)=\frac{g^2}{2\pi\sqrt{\tilde{p}^2}}\frac{\tilde{p}^\mu\tilde{p}^\nu}{\tilde{p}^2}.
\end{eqnarray}
As discussed in the previous section, it would be necessary to insert such a term in higher loop orders, which will cause problems. It might also be mentioned at this point that it is not possible to get rid of this divergence by making the photon massive via a gauge invariant Chern-Simons term of the form
\begin{align}
-\int d^3x\big(&\mu_s\epsilon^{\mu\nu\rho}A_\mu\partial_\nu A_\rho\big)(x),
\end{align}
in the action (\ref{S1}): Expression (\ref{IR1}) stays {\em independent} of the mass parameter $\mu_s$ introduced in such a way. (A similar result is quoted in~\cite{Susskind} for $\Phi^4$-theory.)

However, these problems might be solved by extending the model and introducing the Slavnov term $\frac{\lambda}{2}\star\theta^{\mu\nu}F_{\mu\nu}$ in the action (\ref{S1}), as mentioned previously. The Slavnov extension modifies the graphs shown in Figures~\ref{fig:bosgraphs}b) and \ref{fig:bosgraphs}c) via the modified photon propagator and adds those including the $\lambda$-photon vertex shown in Figures~\ref{fig:bosgraphs}d)-\ref{fig:bosgraphs}f). These new additional terms modify the prefactor of the infrared divergence (\ref{IR1}). The interesting thing is, though, that in $2+1$ dimensions this overall factor is zero: The infrared divergence disappears! The reason for this is that since $\theta_{\mu\nu}$ does not have full rank (see (\ref{theta})), only the terms proportional to $(d-3)$ in (\ref{bosgraphs}) contribute to the IR divergence.

As mentioned in Section~\ref{noncomm-MT} the major effect of the Slavnov term is to make the photon propagator transversal with respect to terms like (\ref{IR1}) and hence making insertions of the dangerous IR divergences (see equation (\ref{IR1})) in higher orders irrelevant. In three dimensions, however, this type of divergence vanishes. Unfortunately, though, a new divergence arises due to $\theta^{\mu\nu}$ not having full rank: At every $\lambda$-photon-photon vertex momenta are contracted with $\theta^{\mu\nu}$ and because of this, the integrands of some terms in the sum of the graphs shown in Figures~\ref{fig:bosgraphs}d)-\ref{fig:bosgraphs}f) become {\em independent} of $k^0$, the zero component of the internal momentum (c.f. second term of the integrand of equation (\ref{bosgraphs})). This problem has already been mentioned at the end of Section~\ref{noncomm-MT}.

In order to compute these terms mentioned above we introduce a momentum cutoff $\Lambda$
\begin{align}\label{cutoff1}
\int\limits_{-\infty}^{+\infty}dk^0\rightarrow\int\limits_{-\Lambda}^{+\Lambda}dk^0=2\Lambda.
\end{align}
At least the spatial part of integral (\ref{bosgraphs}) remains finite and there is no IR pole. We finally arrive at
\begin{align}\label{UVnpl1}
\Pi_{\text{UV-divergent}}^{\mu\nu,\text{np}}(p)&=\Lambda\frac{g^2}{\pi^2}\frac{\tilde{p}^\mu\tilde{p}^\nu}{\tilde{p}^2},
\end{align}
which is a new ultraviolet divergent term of the {\em non-planar} sector: This behaviour is something completely new, since models without the Slavnov term only produce UV divergences in the {\em planar} sector. However, the good news is that (\ref{UVnpl1}) has a structure similar to (\ref{IR1}) and is therefore integrable since it is transversal with respect to our Slavnov-modified photon propagator (see Table~\ref{tab:propMT})\footnote{The renormalization procedure including these new UV divergences will be presented elsewhere \cite{workinprogress}.}. Unfortunately, when examining the planar sector of graphs \ref{fig:bosgraphs}d)-\ref{fig:bosgraphs}f), similar problems arise in performing the $k^0$ integration: A cutoff $\Lambda$ is needed once more and the resulting ultraviolet divergence takes the (transversal) form
\begin{align}\label{UVpl2}
\Pi_{\text{UV-divergent}}^{\mu\nu,\text{pl}}(p)&=\lim_{\Lambda\to\infty}\Lambda\lim_{\epsilon\to0}\Gamma\left(\epsilon\right)\frac{g^2}{2\pi^2\theta^2}\theta^{\mu\r}\left(g_{\r\s}-\frac{\tilde{p}_\r\tilde{p}_\s}{\tilde{p}^2}\right)\theta^{\s\nu},
\end{align}
where $\epsilon\equiv(3-d)$ needed to be introduced due to dimensional regularization needed for the momentum integrations. Also note that the dimension of the coupling constant in (\ref{UVpl2}) is $[g]=(1+\epsilon)/2$.

The new ultraviolet divergence (\ref{UVpl2}) is not proportional to $\tilde{p}^\mu\tilde{p}^\nu$, but fortunately turns out to vanish when acting on our Slavnov-modified photon propagator\footnote{This relation can be easily worked out using (\ref{theta}).}:
\begin{align}\label{uv-null}
&\left(g_{\mu\nu}+(\alpha-1)\frac{p_\mu p_\nu}{p^2}-\frac{\tilde{p}_\mu\tilde{p}_\nu}{\tilde{p}^2}\right)\theta^{\nu\r}\left(g_{\r\s}-\frac{\tilde{p}_\r\tilde{p}_\s}{\tilde{p}^2}\right)\theta^{\s\e}=0.
\end{align}
This result is rather nice because it means that the Slavnov trick still works even though (linear) ultraviolet divergences appear. Also, notice that all linear divergences are independent of the gauge parameter $\alpha$. However, future studies need to check whether logarithmic divergences appear as well, but these would be integrable anyway. Furthermore, we need to stress, that all linear divergences of the photon self-energy are gauge invariant. However, (\ref{UVpl2}) acting on (\ref{photon-prop}) does {\em not} vanish leading to further problems when considering an axial gauge instead of a covariant one.
\section{Non-commutative SQED$_4$}\label{sqed}
\subsection{The Model}
We now consider a coupling of the gauge field $A_{\mu}$ to a complex massive scalar field $\phi$ with a fourth order potential (i.e. \cite{Belov}). The arising theory will be called non-commutative scalar quantum electrodynamics (NCSQED). For reasons which will become evident in this section, we Wick-rotate into flat four dimensional Euclidean space with the metric $g_{\mu\nu}=\delta_{\mu\nu}=\text{diag}(+1,+1,+1,+1)$. In order to make this also visible in our notation, we write all formulas with lower indices. The classical action defining the model using a covariant gauge is thus given by
\begin{align}\label{HLagrangian}
&\Gamma^{(0)}_{\text{SQED}}=\int d^4 x\ \bigg(D_{\mu}\phi^{*}\star D_{\mu}\phi+m^2\phi\star\phi^{*}+\frac{1}{4}F_{\mu\nu}\star F_{\mu\nu}+\frac{1}{2}\lambda\star\theta_{\mu\nu}F_{\mu\nu}-\nonumber\\
&\quad\quad-B\star\partial_{\mu}A_{\mu}-\frac{\alpha}{2}B\star B+\bar{c}\star \partial_{\mu}D_{\mu}c+a\ \phi^{*}\star\phi\star\phi^{*}\star\phi+b\ \phi^{*}\star\phi^{*}\star\phi\star\phi\bigg)(x),
\end{align}
with all fields transforming in the $U(1)$ gauge group and the covariant derivative defined as
\begin{equation}
D_{\mu}\dots :=\partial_{\mu}-ig[A_{\mu}\starco\dots].
\end{equation}
It has been shown in \cite{Belov} that for renormalizeability the coupling constants of the $\phi^4$ potential are constrained to fulfill $a=b$.

The model is characterized by the following BRST transformation rules for the scalar fields
\begin{align}
&s\phi=ig[c\starco\phi],&&s(D_{\mu}\phi)=-ig[D_{\mu}\phi\starco c],\nonumber\\
&s\phi^{*}=ig[c\starco\phi^{*}],&&s(D_{\mu}\phi^{*})=-ig[D_{\mu}\phi^{*}\starco c],\nonumber\\
&s^2\phi=s^2\phi^{*}=0,
\end{align}
and additionally one has also the set of transformations given in (\ref{BRS}).

Before we start with the calculation of loop graphs, we try to find a power counting formula in order to estimate the superficial degree of divergence of a given Feynman graph. The result for the UV region is: 
\begin{equation}\label{Hpowercounting}
d(\gamma)=4-E_A-2E_{\lambda}-E_{\phi}-E_{c}.
\end{equation}
In contrast to (\ref{dgammasl1}) there is no dependence on the coupling, meaning that the superficial degree of divergence is uniquely defined by the number of external legs and is independent of the actual internal structure.

Applying (\ref{Hpowercounting}) to the two point function, we notice that the highest possible value for $d(\gamma)$ is $2$. Especially, the photon self-energy ($E_A=2$) and the scalar self-energy ($E_{\phi}=2$) are predicted to show quadratic UV divergences in the worst case, which may induce quadratic IR singularities as well. This result is indeed verified by explicit calculations at the one-loop level, as we will see later on.
\subsection{One-loop Calculations}
\subsubsection{The Scalar Self-Energy}
For the self-energy, we have to consider the graphs depicted in Figure~\ref{fig:scalfieldcorr}.
\FIGURE[h]{\epsfig{file=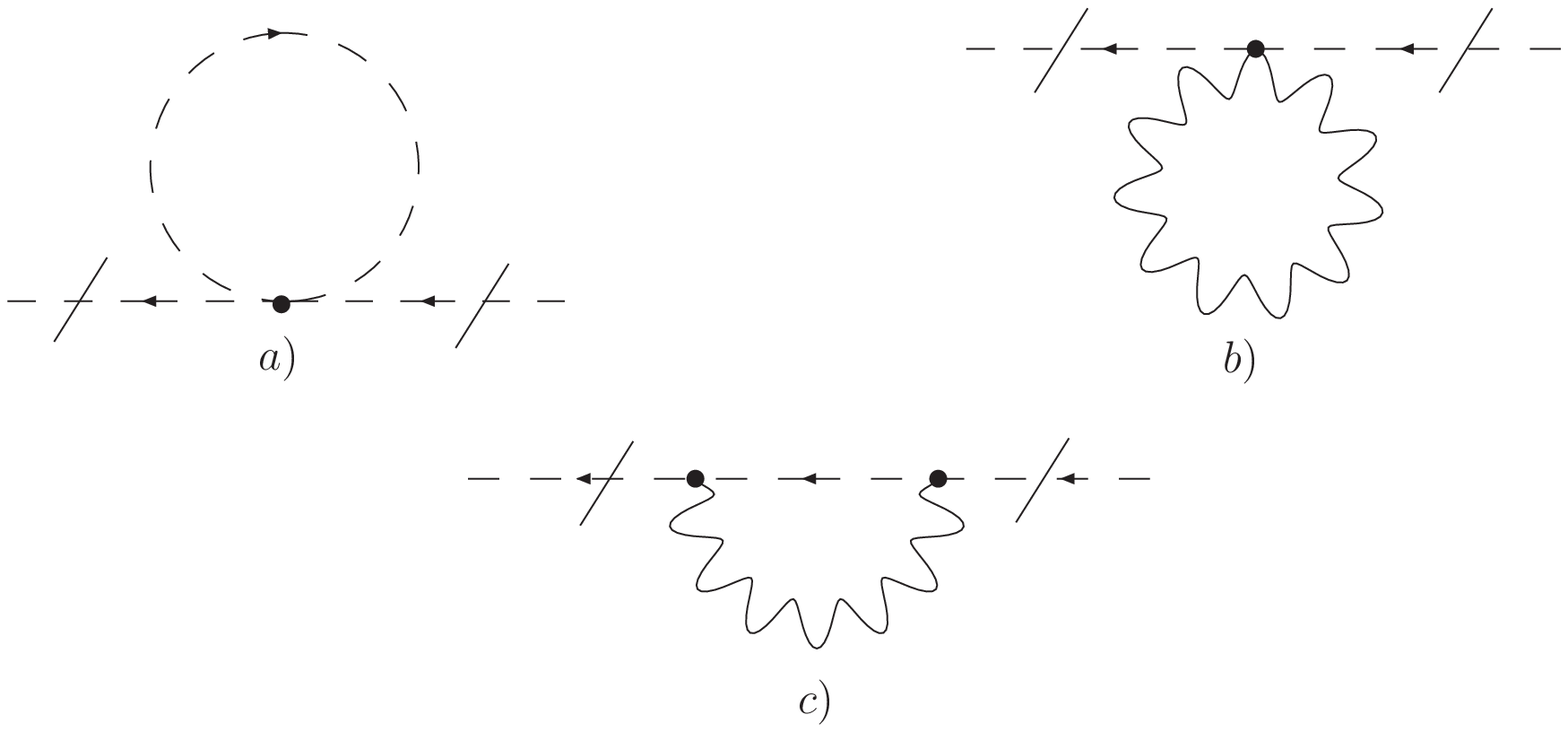, width=10.5cm}
\caption{all possible one-loop corrections for the scalar field $\phi$ (amputated graphs)}
\label{fig:scalfieldcorr}}
In contrast to NCQED with a coupling to fermionic matter the phase factors of the vertices do not cancel each other leaving us with a non-vanishing non-planar part. The planar part shows the well known UV divergences which are already present in the commutative theory and which have to be treated by the usual renormalization procedure. The aim of the present investigation is again the non-planar structure.

The explicit result reads
\begin{align}
\Sigma^{\text{np}}_{\text{IR-divergent}}(p)=\frac{2g^2-b}{2\pi^2\tilde{p}^2}-\frac{(g^2+b)m^2}{8\pi^2}\
\ln(m^2\tilde{p}^2).
\end{align}
In perfect agreement with the power counting formula (\ref{Hpowercounting}) we encounter a quadratic IR divergence in the first term. As already noted in \cite{Belov} it is possible to eliminate this IR divergence by simply setting $b=2g^2$ leaving just a logarithmic IR pole, which is integrable.

It is furthermore a very interesting result that this outcome is independent of the gauge parameter $\alpha$.
\subsubsection{The Photon Self-Energy}
Due to the new $\lambda$ field, we have to deal with the graphs involving the $\lambda$ vertex as depicted in Figure~\ref{fig:bosgraphs}. Additionally, we have two graphs including interactions with the scalar field (see Figure~\ref{fig:bosfieldcorr}).
\FIGURE[h]{\epsfig{file=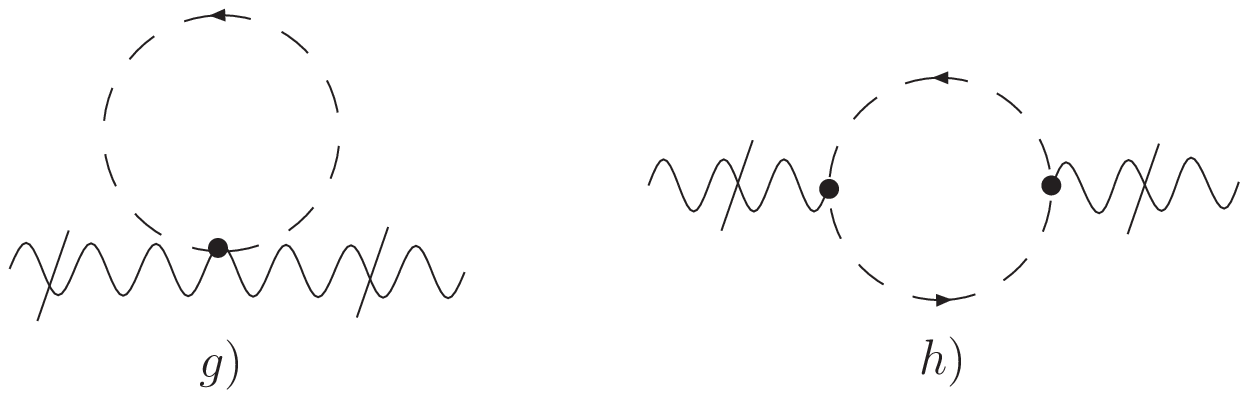, width=8.5cm}
\caption{additional one-loop corrections to the photon propagator --- amputated graphs}
\label{fig:bosfieldcorr}} 
For the technical evaluation of these loop integrals it is essential to specify a certain $\theta_{\mu\nu}$. As a first choice we stick to \cite{Slavnov}, where the following $\theta_{\mu\nu}$ was used
\begin{equation}\label{Htheta1}
\theta_{\mu\nu}=\theta\left(\begin{array}{cccc} 0 & 0 & 0 & 0 \\ 0 & 0 & 1 & 0 \\ 0 & -1 & 0 & 0 \\ 0 & 0 & 0 & 0 \end{array}\right).
\end{equation}
This, however, leads to the same problems we have had in Section~\ref{qed3}, namely that certain Feynman integrals (the ones for the graphs \ref{fig:bosgraphs}d) and \ref{fig:bosgraphs}f)) become independent of the $k_1$ and $k_4$ directions. We can therefore split off the following integral
\begin{equation}
\Pi_{\mu\nu}^{(d,f)}(p)\propto\int_{-\infty}^{\infty}\!dk_1\int_{-\infty}^{\infty}dk_4,
\end{equation}
which we solve by introducing polar coordinates and an appropriate cutoff parameter
\begin{equation}
\int_{-\infty}^{\infty}\!dk_1\int_{-\infty}^{\infty}\!dk_4=\lim_{\Lambda\to\infty}\int_0^{\Lambda}d\rho\rho\int_0^{2\pi}d\varphi=\lim_{\Lambda\to \infty}\Lambda^2\pi,
\end{equation}
The result for the photon self-energy is
\begin{align}\label{selfenergy1}
\Pi^{\text{np}}_{\mu\nu}(p)&=\frac{g^2}{\pi^2}\bigg[3\ \frac{\tilde{p}_{\mu}\tilde{p}_{\nu}}{(\tilde{p}^2)^2}+\frac{\Lambda^2\tilde{p}_{\mu}\tilde{p}_{\nu}}{4\tilde{p}^2}\bigg]+\ldots.
\end{align}
In order to avoid the quadratic UV divergence in (\ref{selfenergy1}) we now choose a $\theta_{\mu\nu}$ with full rank
\begin{equation}\label{Htheta2}
\theta_{\mu\nu}=\left(\begin{array}{cccc} 0 & \theta_1 & 0 & 0 \\ -\theta_1 & 0 & 0 & 0 \\ 0 & 0 & 0 & \theta_2 \\ 0 & 0 & -\theta_2 & 0 \end{array}\right),\hspace{1cm} \theta_1,\theta_2\in\mathbb{R}.
\end{equation}
Going through the same calculations again using this $\theta_{\mu\nu}$, one finds for the self-energy of the gauge boson this time
\begin{align}\label{selfenergy2}
\hat{\Pi}^{\text{np}}_{\mu\nu}(p)=&\frac{g^2}{\pi^2}\bigg[3\ \frac{\tilde{p}_{\mu}\tilde{p}_{\nu}}{(\tilde{p}^2)^2}+\frac{\tilde{p}_{\mu}\tilde{p}_{\nu}}{\theta_1^2\theta_2^2(p^2)^2}\bigg]+\ldots,
\end{align}
showing no more UV divergences.
\section{Higher Loop Orders}\label{higher-order}
We have now shown that the IR divergent terms at one-loop order are
\begin{itemize}
\item proportional to $\Pi_{\text{IR}}=\frac{\tilde{p}_\mu\tilde{p}_\nu}{(\tilde{p}^2)^2}$ and
\item that this structure is independent of couplings to fermions or scalar fields --- only the overall factor is changed when coupling to scalar fields,
\item and these IR singularities are gauge independent~\cite{paper3,ruiz}.
\end{itemize}
Therefore, the graph depicted in Figure~\ref{fig:2-loop}a) is free of non-integrable IR singularities. However, we have also found out that, in case $\theta_{\mu\nu}$ does not have full rank, new UV divergences in both planar as well as non-planar graphs appear. Those coming from non-planar graphs also have the structure $\tilde{p}_\mu\tilde{p}_\nu$ and hence drop out before integrating out Figure~\ref{fig:2-loop}a). UV singularities coming from the planar graphs may still cause problems.
\FIGURE[h]{\includegraphics[scale=0.6]{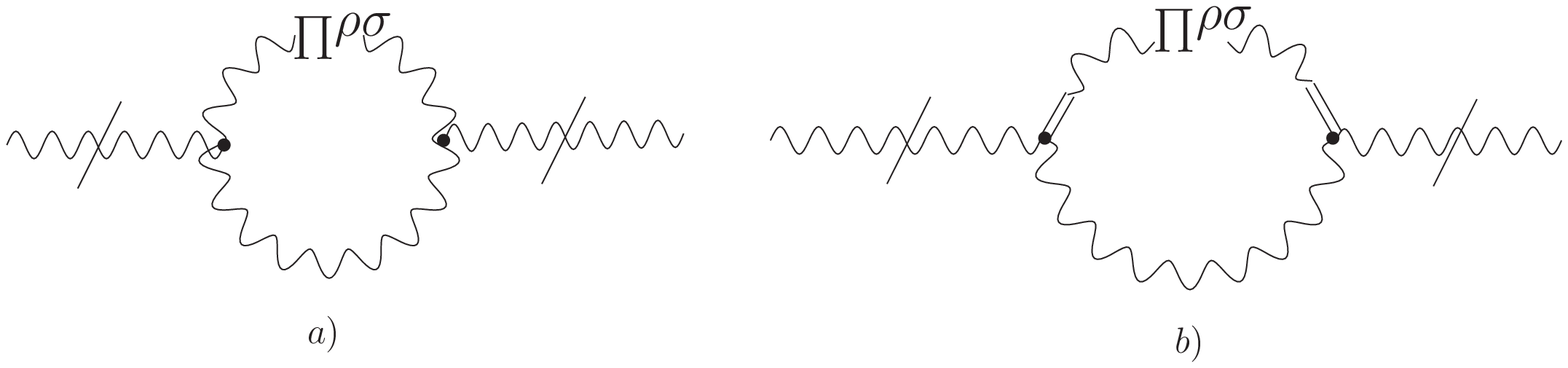}
\caption{some (amputated) 2-loop graphs}
\label{fig:2-loop}}

Unfortunately, due to the existence of the $\lambda$-vertex, the number of graphs is greatly increased and at 2-loop level one can also construct graphs for which Slavnov's trick does not work. An example is depicted in Figure~\ref{fig:2-loop}b). We have computed this graph in 4-dimensional euclidian space (for simplicity in Feynman gauge $\alpha=1$, $\xi=0$) and using (\ref{selfenergy2}) with $\theta_1=\theta_2$ we have found
\begin{align}\label{2l-prob}
\Pi^{2L,IR,b}_{\mu\nu}(p)&=\frac{8g^4}{\pi^2}\int\frac{d^4k}{(2\pi)^4}\sin^2 \left(\frac{k\tilde{p}}{2}\right)\theta_{\mu\r}\left(\delta_{\r\s}-\frac{(\tilde{k}-\tilde{p})_\r(\tilde{k}-\tilde{p})_\s}{(\tilde{k}-\tilde{p})^2}\right)\frac{\theta_{\s\nu}}{(k-p)^2\tilde{k}^4}=\nonumber\\
&=\lim\limits_{\e\to0}\frac{-\ln\e}{16\pi^2\theta^2\tilde{p}^2}\theta_{\r\mu}\left(\delta_{\mu\nu}-\frac{\tilde{p}_\mu\tilde{p}_\nu}{\tilde{p}^2}\right)\theta_{\nu\s}.
\end{align}
This expression seems problematic in two ways: The last parameter integral diverged producing $\lim\limits_{\e\to0}\ln\e$. This was expected since the integral over k-space (\ref{2l-prob}) superficially showed a logarithmic divergence at $k=0$. Second, we observe that $\tilde{p}_\mu\tilde{p}_\nu$ is obviously not the only IR divergent structure in the model: (\ref{2l-prob}) is also transversal with respect to $p_\r$ and shows a quadratic IR divergence as well. Furthermore, letting (\ref{2l-prob}) act on the photon propagator in Feynman gauge is zero, but this need not be true in a more general gauge.

However, there are still many other graphs at 2-loop level which could produce similar results. There are in fact some convincing arguments, why all these problematic terms should cancel: Slavnov, for instance, considered a special axial gauge $n_\mu=(0,1,0,0)$ in~\cite{slavnov2} fixing $A_1=0$. One can then easily work out that, assuming asymptotic boundary conditions, $A_2=0$ follows from Slavnov's constraint $\widetilde{\partial}^\mu A_\mu=0$ when choosing $\theta_{12}=-\theta_{21}=\theta$ as the only non-vanishing components. In this special gauge the relevant term in the action $\lambda\theta^{\mu\nu}[A_\mu\starco A_\nu]$ vanishes. Hence, none of the graphs including the $\lambda$-field exists and none of the problems discussed earlier is present.
\section{Conclusion}
We have proved that the highest order IR divergences in non-commutative U(1) gauge theories including the Slavnov term remain gauge invariant, a result which was expected since it is known from the literature~\cite{paper3,ruiz} that these IR divergences are gauge invariant in theories {\em without} Slavnov's extension. But we have discovered that new UV divergences appear in the model if $\theta_{\mu\nu}$ does not have full rank. These new divergences might present new problems even if the Slavnov term turns out to cure the UV/IR mixing problem.

In Section \ref{qed3} we have found that by extending massless $2+1$ dimensional QED to non-commutative Minkowski space $\mathbb{M}^3_{NC}$ the resulting model is no longer finite but exhibits a linear infrared singularity
\begin{align}\label{concIR1}
\Pi^{\mu\nu}_{IR-divergent}(p)&=\frac{e^2}{2\pi\sqrt{\tilde{p}^2}}\frac{\tilde{p}^\mu\tilde{p}^\nu}{\tilde{p}^2},
\end{align}
in the vacuum polarization. Making the photon massive via a gauge invariant Chern-Simons term does not cure this problem.

A rather surprising result is that adding the Slavnov term in the action (\ref{S1}) not only makes the photon propagator transversal with respect to (\ref{concIR1}) but that the infrared divergence completely vanishes. Instead, linear ultraviolet divergences
\begin{align}\label{concUV1}
\Pi_{UV-divergent}^{\mu\nu}(p)&=\lim\limits_{\Lambda\to\infty}\Lambda\frac{g^2}{\pi^2}\left[\frac{\tilde{p}^\mu\tilde{p}^\nu}{\tilde{p}^2}+\lim_{\epsilon\to0}\Gamma\left(\epsilon\right)\inv{2\theta^2}\theta^{\mu\r}\left(g_{\r\s}-\frac{\tilde{p}_\r\tilde{p}_\s}{\tilde{p}^2}\right)\theta^{\s\nu}\right],
\end{align}
appear (the first term even in the non-planar sector) due to $\theta^{\mu\nu}$ not having full rank as discussed above. However, contraction of (\ref{concUV1}) with the Slavnov-modified propagator of Table~\ref{tab:propMT} is zero as well, and hence the Slavnov trick also works with these ultraviolet divergences as long as a covariant gauge is used.

In contrast to NCQED$_3$, in $4$-dimensional scalar quantum electrodynamics, as discussed in Section~\ref{sqed}, cancellation of the infrared divergence does not take place. But on the other hand it is possible to get rid of the (quadratic) ultraviolet divergences by choosing full rank $\theta_{\mu\nu}$. The worst divergent term is then a quadratic infrared divergence of the form\footnote{cf. (\ref{selfenergy2}) with $\theta_1=\theta_2$}
\begin{align}
\Pi^{\mu\nu}_{IR-divergent}(p)&=4\frac{g^2}{\pi^2}\ \frac{\tilde{p}_{\mu}\tilde{p}_{\nu}}{(\tilde{p}^2)^2}.
\end{align}
Finally, in Section~\ref{higher-order} we discussed graphs at 2-loop level for which Slavnov's trick does not work. However, we argued, that problematic terms coming from these graphs should actually cancel in the sum of all 2-loop graphs since one can find a special axial gauge in which these problematic graphs are no longer present~\cite{slavnov2}.
\acknowledgments
\addcontentsline{toc}{section}{\ \ \ \ \ Acknowledgments}
We would like to thank A.~Bichl, S.~Denk, F.~Gieres, V.~Putz, R.~Stora, M.~Wohlgenannt and R.~Wulkenhaar for helpful discussions. This work was supported by "Fonds zur F\"orderung der Wissenschaftlichen Forschung" (FWF) under contract P15463-N08.
\vspace{5cm}
\begin{appendix}
\section{Feynman Rules in covariant gauge}\label{app-feyn}
%
\begin{center}
\TABULAR[h]{|lcl|}{\hline
& & \\
fermion propagator  & \parbox{3.5cm}{\epsfig{file=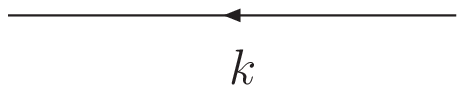, width=3cm}} & \parbox{4cm}{$i\Delta^{\psi\bar{\psi}}(k)=-i\frac{\slashed{k}-m}{k^2-m^2}$}\\[20pt]
photon propagator  & \parbox{3.5cm}{\epsfig{file=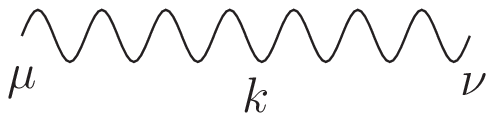, width=3.5cm}} & \parbox{7cm}{$i\Delta^{A}_{\mu\nu}(k)=\frac{-i}{k^2}\left(g_{\mu\nu}-(1-\alpha)\frac{k_{\mu}k_{\nu}}{k^2}-\frac{\tilde{k}_{\mu}\tilde{k}_{\nu}}{\tilde{k}^2}\right)$}\\[20pt]
ghost propagator  & \parbox{3.5cm}{\epsfig{file=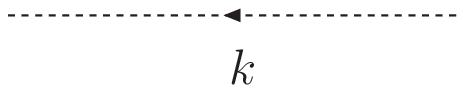, width=3cm}} & \parbox{3cm}{$i\Delta^{c\bar{c}}(k)=\frac{i}{k^2}$}\\[20pt]
$\lambda$ propagator  & \parbox{3.5cm}{\epsfig{file=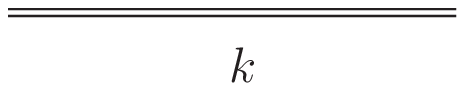, width=3cm}} & \parbox{3cm}{$i\Delta^{\lambda\lambda}(k)=\frac{ik^2}{\tilde{k}^2}$}\\[20pt]
mixed propagator  & \parbox{3.5cm}{\epsfig{file=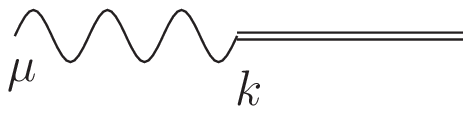, width=3cm}} & \parbox{3cm}{$i\Delta^{\lambda A}_{\mu}(k)=\frac{\tilde{k}_{\mu}}{\tilde{k}^2}$}\\[0.5cm]\hline}{The propagators in $\mathbb{M}^d_{NC}$\label{tab:propMT}}
\TABULAR[h]{|lcl|}{\hline
& & \\
scalar propagator  & \parbox{3.5cm}{\epsfig{file=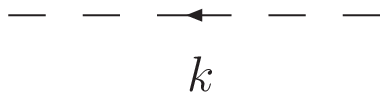, width=3cm}} & \parbox{5cm}{$\Delta^{\phi\phi^{*}}(k)=\frac{1}{k^2+m^2}$}\\[20pt]
photon propagator  & \parbox{3.5cm}{\epsfig{file=photonprop, width=3.5cm}} & \parbox{7cm}{$\Delta^{A}_{\mu\nu}(k)=\inv{k^2}\left(\delta_{\mu\nu}-(1-\alpha)\frac{k_{\mu}k_{\nu}}{k^2}-\frac{\tilde{k}_{\mu}\tilde{k}_{\nu}}{\tilde{k}^2}\right)$}\\[20pt]
ghost propagator  & \parbox{3.5cm}{\epsfig{file=Nghostprop, width=3cm}} & \parbox{5cm}{$\Delta^{c\bar{c}}(k)=-\frac{1}{k^2}$}\\[20pt]
$\lambda$ propagator  & \parbox{3.5cm}{\epsfig{file=lambdalprop, width=3cm}} & \parbox{5cm}{$\Delta^{\lambda\lambda}(k)=-\frac{k^2}{\tilde{k}^2}$}\\[20pt]
mixed propagator  & \parbox{3.5cm}{\epsfig{file=lambdaAprop, width=3cm}} & \parbox{5cm}{$\Delta^{\lambda A}_{\mu}(k)=i\ \frac{\tilde{k}_{\mu}}{\tilde{k}^2}$}\\[0.5cm]\hline}{The propagators in $\mathbb{R}^d_{NC}$\label{tab:prop}}
\TABULAR[h]{ll}{%
\parbox{3cm}{\epsfig{file=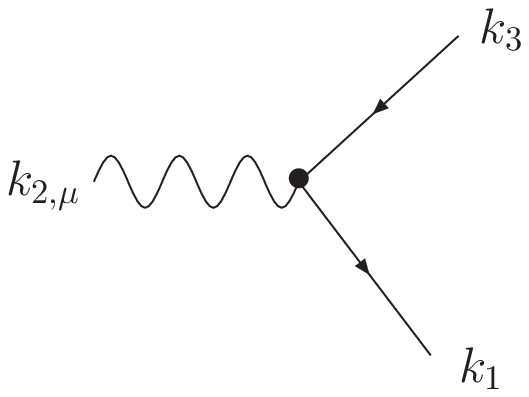, width=2.9cm}} & \parbox{9cm}{$\widetilde{V}^{\bar{\psi}A\psi}_\mu(k_1, k_2, k_3)=ig\gamma_\mu e^{-\frac{i}{2}(k_1\times k_3)}$}\\[1.2cm]
\parbox{3cm}{\epsfig{file=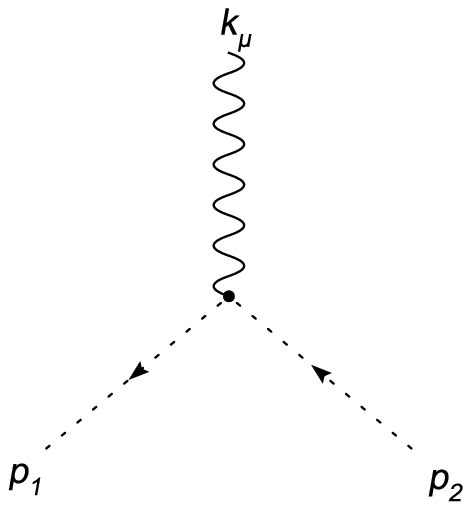, width=2.3cm}} & \parbox{9cm}{$\widetilde{V}^{c\bar{c}A}_{\mu}=2gp_{1,\mu}\sin\left(\frac{p_1\times p_2}{2}\right)$}\\[1.2cm]
\parbox{3cm}{\epsfig{file=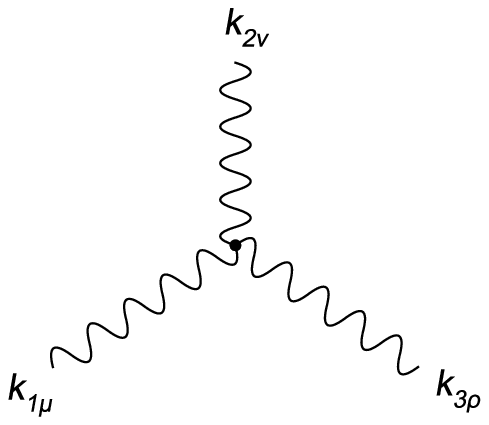, width=2.4cm}} & \parbox{7.2cm}{\begin{align}\widetilde{V}^{AAA}_{\mu\nu\rho}&=2g\sin\left(\frac{k_1\times k_2}{2}\right)\big(g_{\nu\rho}(k_2-k_3)_{\mu}+\nonumber\\
&\quad+g_{\mu\rho}(k_3-k_1)_{\nu}+g_{\mu\nu}(k_1-k_2)_{\rho}\big)\nonumber\end{align}}\\[1.2cm]
\parbox{3cm}{\epsfig{file=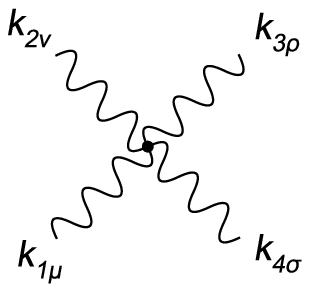, width=2.3cm}} & \parbox{11.15cm}{\begin{align}\widetilde{V}^{AAAA}_{\mu\nu\rho\sigma}&=-4ig^2\bigg[(g_{\mu\rho}g_{\nu\sigma}-g_{\mu\sigma}g_{\nu\rho})\sin\left(\frac{k_1\times k_2}{2}\right)\sin\left(\frac{k_3\times k_4}{2}\right)+\nonumber\\
&\quad\hspace{0.85cm}+(g_{\mu\nu}g_{\rho\sigma}-g_{\mu\sigma}g_{\nu\rho})\sin\left(\frac{k_1\times k_3}{2}\right)\sin\left(\frac{k_2\times k_4}{2}\right)+\nonumber\\
&\quad\hspace{0.85cm}+(g_{\mu\nu}g_{\rho\sigma}-g_{\mu\rho}g_{\nu\sigma})\sin\left(\frac{k_1\times k_4}{2}\right)\sin\left(\frac{k_2\times k_3}{2}\right)\bigg]\nonumber\end{align}}\\[1.2cm]
\parbox{3cm}{\epsfig{file=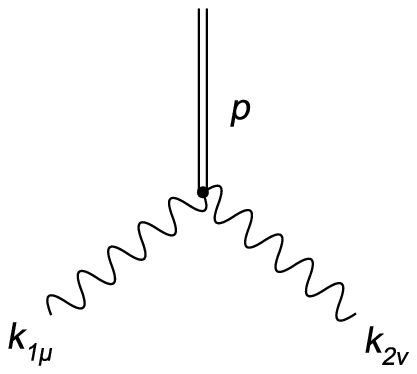, width=2.3cm}} & \parbox{9cm}{$\widetilde{V}_{\mu\nu}^{\lambda AA}=2ig\theta_{\mu\nu}\sin\left(\frac{k_1\times k_2}{2}\right)$}\vspace{0.5cm}}{The vertices in $\mathbb{M}^d_{NC}$\label{tab:vertexMT}}
\TABULAR[h]{ll}{%
\parbox{3cm}{\epsfig{file=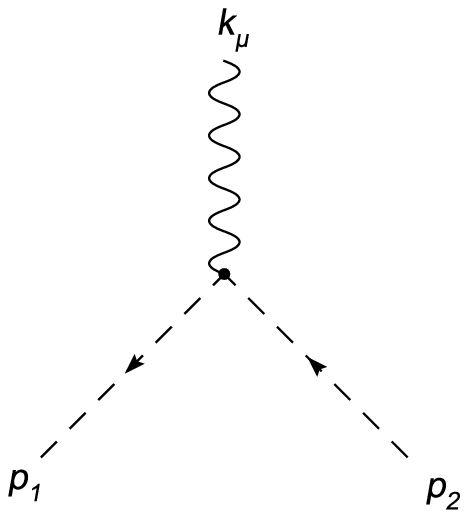, width=1.7cm}} & \parbox{9cm}{$\widetilde{V}_{\mu}^{A\phi\phi^{*}}(k,p_1,p_2)=2ig(p_{1}-p_2)_{\mu}\sin\left(\frac{p_1\times p_2}{2}\right)$}\\[1.2cm]
\parbox{3cm}{\epsfig{file=ghostvertex, width=1.7cm}} & \parbox{9cm}{$\widetilde{V}^{c\bar{c}A}_{\mu}=-2igp_{1,\mu}\sin\left(\frac{p_1\times p_2}{2}\right)$}\\[1.2cm]
\parbox{3cm}{\epsfig{file=photon3vertex, width=1.9cm}} & \parbox{7.55cm}{\begin{align}{\widetilde{V}^{AAA}_{\mu\nu\rho}}&=-2ig\sin\left(\frac{k_1\times k_2}{2}\right)\big(\delta_{\nu\rho}(k_2-k_3)_{\mu}+\nonumber\\
&\quad+\delta_{\mu\rho}(k_3-k_1)_{\nu}+\delta_{\mu\nu}(k_1-k_2)_{\rho}\big)\nonumber\end{align}}\\[1.2cm]
\parbox{3cm}{\epsfig{file=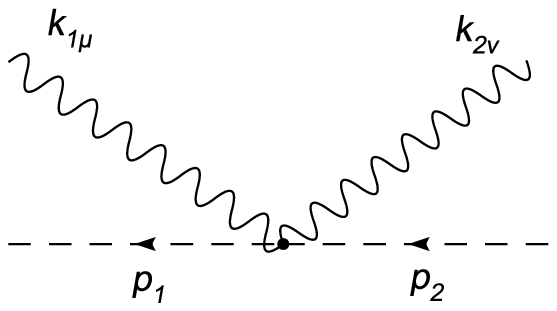, width=1.9cm}} & \parbox{8.75cm}{\begin{align}{\widetilde{V}^{AA\phi\phi^{*}}_{\mu\nu}}=4g^2\delta_{\mu\nu}\bigg[&\cos\left(\frac{k_1\times p_1}{2}+\frac{k_2\times p_2}{2}\right)-\nonumber\\
&-\cos\left(\frac{p_1\times p_2}{2}\right)\cos\left(\frac{k_1\times k_2}{2}\right)\bigg]\nonumber\end{align}}\\[1.2cm]
\parbox{3cm}{\epsfig{file=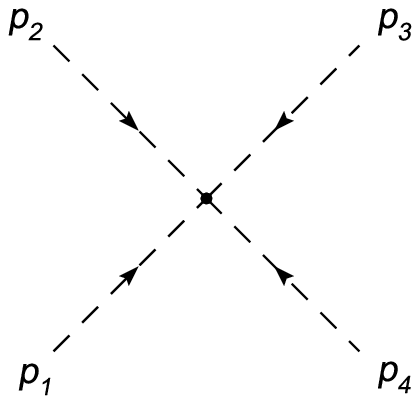, width=1.9cm}} & \parbox{8.45cm}{\begin{align}\widetilde{V}^{\phi\phi^{*}\phi\phi^{*}}=-4\bigg[&a\cos\left(\frac{p_1\times p_2}{2}+\frac{p_3\times p_4}{2}\right)+\nonumber\\
&+b\cos\left(\frac{p_1\times p_3}{2}\right)\cos\left(\frac{p_2\times p_4}{2}\right)\bigg]\nonumber\end{align}}\\[1.2cm]
\parbox{2.2cm}{\epsfig{file=photonvertex, width=2.2cm}} & \parbox{11cm}{\begin{align}\widetilde{V}^{AAAA}_{\mu\nu\rho\sigma}=-4&g^2\bigg[(\delta_{\mu\rho}\delta_{\nu\sigma}-\delta_{\mu\sigma}\delta_{\nu\rho})\sin\left(\frac{k_1\times k_2}{2}\right)\sin\left(\frac{k_3\times k_4}{2}\right)+\nonumber\\
&\ +(\delta_{\mu\nu}\delta_{\rho\sigma}-\delta_{\mu\sigma}\delta_{\nu\rho})\sin\left(\frac{k_1\times k_3}{2}\right)\sin\left(\frac{k_2\times k_4}{2}\right)+\nonumber\\
&\ +(\delta_{\mu\nu}\delta_{\rho\sigma}-\delta_{\mu\rho}\delta_{\nu\sigma})\sin\left(\frac{k_1\times k_4}{2}\right)\sin\left(\frac{k_2\times k_3}{2}\right)\bigg]\nonumber\end{align}}\\[1.2cm]
\parbox{3cm}{\epsfig{file=lambdavertex, width=1.9cm}} & \parbox{9cm}{$\widetilde{V}_{\mu\nu}^{\lambda AA}=2g\theta_{\mu\nu}\sin\left(\frac{k_1\times k_2}{2}\right)$}\vspace{0.5cm}}{The vertices in $\mathbb{R}^d_{NC}$\label{tab:vertex}}
\end{center}
\clearpage
\section{One-loop graphs of the gauge boson self-energy}\label{app-graphs}
The gauge boson self-energy at one-loop level consists of six graphs depicted in Figure~\ref{fig:bosgraphs}: The ghost loop $\Pi^{\mu\nu}_a(p)$ (Fig.~\ref{fig:bosgraphs}a), the tadpole graph $\Pi^{\mu\nu}_b(p)$ (Fig.~\ref{fig:bosgraphs}b), the boson loop $\Pi^{\mu\nu}_c(p)$ (Fig.~\ref{fig:bosgraphs}c), the graph with one inner $\lambda$-propagator $\Pi^{\mu\nu}_d(p)$ (Fig.~\ref{fig:bosgraphs}d), the graph with one inner $\lambda A$-propagator $\Pi^{\mu\nu}_e(p)$ (Fig.~\ref{fig:bosgraphs}e) and the graph with two inner $\lambda A$-propagators $\Pi^{\mu\nu}_f(p)$ (Fig.~\ref{fig:bosgraphs}f). The first term in the expansion (\ref{expansion}) is then given by
\begin{align}
\int d^dk I^{\mu \nu} (k, 0)& \sin^2 \left(\frac{k\tilde{p}}{2}\right) = \int d^dk \sum\limits_{i=a-f}I^{\mu \nu}_i (k, 0) \sin^2 \left(\frac{k\tilde{p}}{2}\right) \equiv\nonumber\\
&\quad\equiv i\Pi^{\mu\nu}_{a,IR}(p)+i\Pi^{\mu\nu}_{b,IR}(p)+i\Pi^{\mu\nu}_{c,IR}(p)+i\Pi^{\mu\nu}_{d,IR}(p)+i\Pi^{\mu\nu}_{e,IR}(p)+i\Pi^{\mu\nu}_{f,IR}(p).
\end{align}
\subsection*{Ghost-loop:}
The ghost propagator and vertex in interpolating gauge are given by
\begin{align}
\Delta^{c\bar{c}}(k)&=\frac{i}{k^2-\zeta(nk)^2},\nonumber\\
V^{c\bar{c}A}_\mu(q_1, q_2, k)&=2g\left(q_{2\mu}-\zeta(nq_2)n_\mu\right)\sin\left(\frac{q_1\tilde{q}_2}{2}\right),
\end{align}
where $\zeta=\xi/n^2$. Therefore one gets for the ghost loop graph depicted in Figure~\ref{fig:bosgraphs}a)
\begin{align}
i\Pi^{\mu\nu}_{a,IR}(p)&=4g^2\int\frac{d^dk}{(2\pi)^d}\sin^2\left(\frac{k\tilde{p}}{2}\right)\frac{-\left[k^\mu-\zeta(nk)n^\mu\right]\left[k^\nu-\zeta(nk)n^\nu\right]}{\left[k^2-\zeta(nk)^2\right]^2}=\nonumber\\
&=4g^2\int\frac{d^dk}{(2\pi)^d}\sin^2\left(\frac{k\tilde{p}}{2}\right)\big\{\frac{-k^\mu k^\nu}{\left[k^2-\zeta(nk)^2\right]^2}+b\frac{\left(n^\mu k^\nu+k^\mu n^\nu\right)}{k^2-\zeta(nk)^2}-b^2n^\mu n^\nu\big\}
\end{align}
\subsection*{Tadpole:}
With the boson propagator given in (\ref{photon-prop}) and the $4A$-vertex given in Table~\ref{tab:vertexMT} one gets for the graph depicted in Figure~\ref{fig:bosgraphs}b)
\begin{align}
i\Pi^{\mu\nu}_{b,IR}(p)&=2g^2\int\frac{d^dk}{(2\pi)^d}\sin^2\left(\frac{k\tilde{p}}{2}\right)\inv{k^2}\left(g^{\mu\tau}g^{\s\nu}+g^{\mu\s}g^{\tau\nu}-2g^{\mu\nu}g^{\s\tau}\right)\times\nonumber\\
&\quad\qquad\qquad\times\left[g_{\t\s}-ak_\t k_\s+b(n_\t k_\s+k_\t n_\s)-b'\left(\tilde{k}_\t k_\s+k_\t\tilde{k}_\s\right)-\frac{\tilde{k}_\t\tilde{k}_\s}{\tilde{k}^2}\right]\nonumber\\
&=4g^2\int\frac{d^dk}{(2\pi)^d}\sin^2\left(\frac{k\tilde{p}}{2}\right)\inv{k^2}\big\{g^{\mu\nu}\left[k^2a-d+2-2(nk)b\right]+b\left(n^\mu k^\nu+k^\mu n^\nu\right)-\nonumber\\
&\quad\qquad\qquad-ak^\mu k^\nu-b'\left(\tilde{k}^\mu k^\nu+k^\mu\tilde{k}^\nu\right)-\frac{\tilde{k}^\mu\tilde{k}^\nu}{\tilde{k}^2}\big\},
\end{align}
where $d=g^\mu_{\ \mu}$ denotes the dimension of space-time and $a$, $b$, and $b'$ were defined in (\ref{defabb'}).
\subsection*{Photon-loop:}
Consulting the Feynman rules given in Table~\ref{tab:vertexMT} and equation (\ref{photon-prop}) one has for the graph depicted in Figure~\ref{fig:bosgraphs}c)
\begin{align}
i\Pi^{\mu\nu}_{c,IR}(p)&=2g^2\int\frac{d^dk}{(2\pi)^d}\inv{k^4}\left[-k^\e g^{\mu\s}+2k^\mu g^{\e\s}-k^\s g^{\e\mu}\right]\left[-k^\r g^{\nu\t}+2k^\nu g^{\r\t}-k^\t g^{\r\nu}\right]\times\nonumber\\
&\quad\times\sin^2\left(\frac{k\tilde{p}}{2}\right)\left[g_{\t\e}-ak_\t k_\e+b(n_\t k_\e+k_\t n_\e)-b'\left(\tilde{k}_\t k_\e+k_\t\tilde{k}_\e\right)-\frac{\tilde{k}_\t\tilde{k}_\e}{\tilde{k}^2}\right]\times\nonumber\\
&\quad\times\left[g_{\s\r}-ak_\s k_\r+b(n_\s k_\r+k_\s n_\r)-b'\left(\tilde{k}_\s k_\r+k_\s\tilde{k}_\r\right)-\frac{\tilde{k}_\s\tilde{k}_\r}{\tilde{k}^2}\right].
\end{align}
Noticing that
\begin{align}\label{nr1}
&\left[-k^\e g^{\mu\s}+2k^\mu g^{\e\s}-k^\s g^{\e\mu}\right]\left[g_\t^{\ \e}-ak_\t k_\e+b(n_\t k_\e+k_\t n_\e)-b'\left(\tilde{k}_\t k_\e+k_\t\tilde{k}_\e\right)-\frac{\tilde{k}_\t\tilde{k}_\e}{\tilde{k}^2}\right]=\nonumber\\
&=\Big[-k_\t g^{\mu\s}+2k^\mu g_\t^{\ \s}-k^\s g_\t^{\ \mu}+ak_\t(k^2g^{\mu\s}-k^\mu k^\s)+bn_\t(k^\mu k^\s-k^2g^{\mu\s})+bk_\t(-nkg^{\mu\s}+\nonumber\\
&\;\:+2k^\mu n^\s-n^\mu k^\s)+b'k^2\tilde{k}_\t g^{\mu\s}-b'\tilde{k}_\t k^\mu k^\s-2b'\tilde{k}^\s k_\t k^\mu+b'\tilde{k}^\mu k_\t k^\s-2k^\mu\frac{\tilde{k}_\t\tilde{k}^\s}{\tilde{k}^2}+k^\s\frac{\tilde{k}_\t\tilde{k}^\mu}{\tilde{k}^2}\Big]\nonumber\\
&=\Big[fk_\t g^{\mu\s}-k_\t k^\s(ak^\mu+bn^\mu-b'\tilde{k}^\mu)+2k^\mu g_\t^{\ \s}-k^\s g_\t^{\ \mu}+(k^\mu k^\s-k^2g^{\mu\s})(bn_\t-b'\tilde{k}_\t)+\nonumber\\
&\;\:+2bk^\mu k_\t n^\s-2b'\tilde{k}^\s k_\t k^\mu-2k^\mu\frac{\tilde{k}_\t\tilde{k}^\s}{\tilde{k}^2}+k^\s\frac{\tilde{k}_\t\tilde{k}^\mu}{\tilde{k}^2}\Big],
\end{align}
with the abbreviation
\begin{align}\label{deff}
f=k^2a-1-(nk)b,
\end{align}
we get
\begin{align}
i\Pi^{\mu\nu}_{c,IR}(p)&=2g^2\int\frac{d^dk}{(2\pi)^d}\sin^2\left(\frac{k\tilde{p}}{2}\right)\Bigg[fk_\t g^{\mu\s}-k_\t k^\s(ak^\mu+bn^\mu-b'\tilde{k}^\mu)+2k^\mu g_\t^{\ \s}-k^\s g_\t^{\ \mu}+\nonumber\\
&\;\,+2bk^\mu k_\t n^\s+(k^\mu k^\s-k^2g^{\mu\s})(bn_\t-b'\tilde{k}_\t)-2b'\tilde{k}^\s k_\t k^\mu-2k^\mu\frac{\tilde{k}_\t\tilde{k}^\s}{\tilde{k}^2}+k^\s\frac{\tilde{k}_\t\tilde{k}^\mu}{\tilde{k}^2}\Bigg]\times\nonumber\\
&\;\,\times\Bigg[fk_\s g^{\nu\t}-k_\s k^\t(ak^\nu+bn^\nu-b'\tilde{k}^\nu)+2k^\nu g_\s^{\ \t}-k^\t g_\s^{\ \nu}+2bk^\nu k_\s n^\t+\nonumber\\
&\;\,+(k^\nu k^\t-k^2g^{\nu\t})(bn_\s-b'\tilde{k}_\s)-2b'\tilde{k}^\t k_\s k^\nu-2k^\nu\frac{\tilde{k}_\s\tilde{k}^\t}{\tilde{k}^2}+k^\t\frac{\tilde{k}_\s\tilde{k}^\nu}{\tilde{k}^2}\Bigg]\inv{k^4},
\end{align}
leading to
\begin{align}
i\Pi^{\mu\nu}_{c,IR}(p)&=2g^2\int\frac{d^dk}{(2\pi)^d}\sin^2\left(\frac{k\tilde{p}}{2}\right)\inv{k^2}\bigg\{2k^2b^2n^\mu n^\nu+2g^{\mu\nu}\left[(nk)b-f\right]+\nonumber\\
&\quad+\frac{k^\mu k^\nu}{k^2}\big[f^2-2k^2af+4f+4(nk)bf+k^4a^2-2k^2a-4k^2(nk)ab+4d-\nonumber\\
&\quad-7+10(nk)b+5(nk)^2b^2\big]+2\frac{\tilde{k}^\mu\tilde{k}^\nu}{\tilde{k}^2}\left[f-(nk)b+k^2\tilde{k}^2b'^2\right]+\nonumber\\
&\quad+b\left(n^\mu k^\nu+k^\mu n^\nu\right)\left[k^2a-f-5-3(nk)b\right]-2k^2b'b\left(\tilde{k}^\mu n^\nu+n^\mu\tilde{k}^\nu\right)+\nonumber\\
&\quad+b'\left(\tilde{k}^\mu k^\nu+k^\mu\tilde{k}^\nu\right)\left[f-k^2a+5+3(nk)b\right]\bigg\},
\end{align}
where $d=g^\mu_{\ \mu}$ once more. Using (\ref{defabb'}) and (\ref{deff}) this expression becomes
\begin{align}
i\Pi^{\mu\nu}_{c,IR}(p)&=4g^2\int\frac{d^dk}{(2\pi)^d}\sin^2\left(\frac{k\tilde{p}}{2}\right)\inv{k^2}\bigg\{k^2b^2n^\mu n^\nu-g^{\mu\nu}\left[k^2a-1-2(nk)b\right]-\nonumber\\
&\quad+\frac{k^\mu k^\nu}{k^2}\left[k^2a+2d-5+2(nk)b+(nk)^2b^2\right]-b\left(n^\mu k^\nu+k^\mu n^\nu\right)\left[2+(nk)b\right]+\nonumber\\
&\quad+\frac{\tilde{k}^\mu\tilde{k}^\nu}{\tilde{k}^2}\left(k^2a-1-2(nk)b+k^2\tilde{k}^2b'^2\right)-k^2b'b(\tilde{k}^\mu n^\nu+n^\mu\tilde{k}^\nu)+\nonumber\\
&\quad+b'(\tilde{k}^\mu k^\nu+k^\mu\tilde{k}^\nu)\left[2+(nk)b\right]\bigg\}.
\end{align}
\subsection*{Graph with inner $\lambda$-propagator:}
With the boson propagator (\ref{photon-prop}) and the $\lambda$-propagator and vertex given in Table~\ref{tab:propMT} one gets for the graph depicted in Figure~\ref{fig:bosgraphs}d)
\begin{align}
i\Pi^{\mu\nu}_{d,IR}(p)=4g^2\int\frac{d^dk}{(2\pi)^d}\sin^2\left(\frac{k\tilde{p}}{2}\right)\inv{\tilde{k}^2}\theta^{\mu\t}&\Bigg[g_{\t\s}-ak_\t k_\s+b(n_\t k_\s+k_\t n_\s)-\nonumber\\
&\quad-b'\left(\tilde{k}_\t k_\s+k_\t\tilde{k}_\s\right)-\frac{\tilde{k}_\t\tilde{k}_\s}{\tilde{k}^2}\Bigg]\theta^{\s\nu}.
\end{align}
\subsection*{Graph with one inner $\lambda$-A propagator:}
The mixed $\lambda$-A propagator in interpolating gauge is given by
\begin{align}
\Delta^{\lambda A}_\mu(k)=\frac{\tilde{k}_\mu}{\tilde{k}^2}+b'k_\mu.
\end{align}
The other Feynman rules needed for the graph depicted in Figure~\ref{fig:bosgraphs}e) are given in Table~\ref{tab:vertexMT} and (\ref{photon-prop}). Additionally there is also a graph with an inner A-$\lambda$ propagator instead of a $\lambda$-A propagator, but this graph only corresponds to exchanging the external indices. In the following these additional terms will be abbreviated with '$+\mu\leftrightarrow\nu$'. One gets
\begin{align}
i\Pi^{\mu\nu}_{e,IR}(p)&=4g^2\int\frac{d^dk}{(2\pi)^d}\sin^2\left(\frac{k\tilde{p}}{2}\right)\frac{\theta^{\mu\t}}{k^2\tilde{k}^2}\left(\tilde{k}_\rho+\tilde{k}^2b'k_\rho\right)\left[k^\r g^{\nu\s}-2k^\nu g^{\r\s}+k^\s g^{\r\nu}\right]\times\nonumber\\
&\quad\times\left[g_{\t\s}-ak_\t k_\s+b(n_\t k_\s+k_\t n_\s)-b'\left(\tilde{k}_\t k_\s+k_\t\tilde{k}_\s\right)-\frac{\tilde{k}_\t\tilde{k}_\s}{\tilde{k}^2}\right]+\mu\leftrightarrow\nu
\end{align}
Using (\ref{nr1}) this expression becomes
\begin{align}
i\Pi^{\mu\nu}_{e,IR}(p)&=4g^2\int\frac{d^dk}{(2\pi)^d}\sin^2\left(\frac{k\tilde{p}}{2}\right)\frac{\theta^{\mu\t}}{k^2\tilde{k}^2}\Bigg\{k^2\left(b\tilde{k}^\nu n_\t-2b'\tilde{k}^\nu\tilde{k}_\t+\tilde{k}^2bb'k_\t n^\nu+\tilde{k}^2b'g^\nu_{\ \t}\right)-\nonumber\\
&\quad-\left[f+k^2\tilde{k}^2b'^2\right]k_\t\tilde{k}^\nu-\left[2(n\tilde{k})b+\tilde{k}^2b'\left(f-k^2a+2(nk)b\right)\right]k^\nu k_\t\Bigg\}+\mu\leftrightarrow\nu=\nonumber\\
&=4g^2\int\frac{d^dk}{(2\pi)^d}\sin^2\left(\frac{k\tilde{p}}{2}\right)\frac{-1}{k^2\tilde{k}^2}\Bigg\{2\left[f+k^2\tilde{k}^2b'^2\right]\tilde{k}^\mu\tilde{k}^\nu-k^2\tilde{k}^2bb'(\tilde{k}^\mu n^\nu+ n^\mu\tilde{k}^\nu)+\nonumber\\
&\quad+\left[2(n\tilde{k})b+\tilde{k}^2b'\left(f-k^2a+2(nk)b\right)\right](\tilde{k}^\mu k^\nu+k^\mu\tilde{k}^\nu)-k^2b(\tilde{k}^\mu \tilde{n}^\nu+\tilde{n}^\mu\tilde{k}^\nu)+\nonumber\\
&\quad+2k^2b'(\theta^{\mu\t}\tilde{k}_\t\tilde{k}^\nu+\tilde{k}^\mu\theta^{\nu\t}\tilde{k}_\t)\Bigg\},
\end{align}
and inserting (\ref{defabb'}) and (\ref{deff}) finally leads to
\begin{align}
i\Pi^{\mu\nu}_{e,IR}(p)&=4g^2\int\frac{d^dk}{(2\pi)^d}\sin^2\left(\frac{k\tilde{p}}{2}\right)\frac{1}{k^2\tilde{k}^2}\Bigg\{2\left[1+(nk)b-k^2a-k^2\tilde{k}^2b'^2\right]\tilde{k}^\mu\tilde{k}^\nu+\nonumber\\
&\quad+k^2\tilde{k}^2bb'(\tilde{k}^\mu n^\nu+ n^\mu\tilde{k}^\nu)-\tilde{k}^2b'\left[1+(nk)b\right](\tilde{k}^\mu k^\nu+k^\mu\tilde{k}^\nu)+\nonumber\\
&\quad+k^2b(\tilde{k}^\mu \tilde{n}^\nu+\tilde{n}^\mu\tilde{k}^\nu)-2k^2b'(\theta^{\mu\t}\tilde{k}_\t\tilde{k}^\nu+\tilde{k}^\mu\theta^{\nu\t}\tilde{k}_\t)\Bigg\}.
\end{align}
\subsection*{Graph with two inner $\lambda$-A propagators:}
For the graph depicted in Figure~\ref{fig:bosgraphs}f) we get
\begin{align}
i\Pi^{\mu\nu}_{f,IR}(p)&=4g^2\int\frac{d^dk}{(2\pi)^d}\sin^2\left(\frac{k\tilde{p}}{2}\right)\inv{\tilde{k}^4}\theta^{\mu\t}\theta^{\nu\s}\left(\tilde{k}_\t+\tilde{k}^2b'k_\t\right)\left(\tilde{k}_\s+\tilde{k}^2b'k_\s\right)=\nonumber\\
&=4g^2\int\frac{d^dk}{(2\pi)^d}\sin^2\left(\frac{k\tilde{p}}{2}\right)\inv{\tilde{k}^4}\theta^{\mu\t}\left(-\tilde{k}_\t\tilde{k}_\s-\tilde{k}^2b'(\tilde{k}_\t k_\s+k_\t\tilde{k}_\s)-\tilde{k}^4b'^2k_\t k_\s\right)\theta^{\s\nu}.
\end{align}
\subsection*{The sum of all six graphs is given by:}
(considering $k_\s\theta^{\s\nu}=-\tilde{k}^\nu$)
\begin{align}
i\Pi^{\mu\nu}_{IR}(p)&=4g^2\int\frac{d^dk}{(2\pi)^d}\sin^2\left(\frac{k\tilde{p}}{2}\right)\bigg\{-\frac{b}{k^2}\left(n^\mu k^\nu+k^\mu n^\nu\right)\left[1+(nk)b-\frac{k^2}{k^2-\zeta(nk)^2}\right]-\nonumber\\
&\quad\qquad-\frac{(d-3)}{k^2}g^{\mu\nu}+\frac{k^\mu k^\nu}{k^4}\left[2d-5+2(nk)b+(nk)^2b^2-\frac{k^4}{\left[k^2-\zeta(nk)^2\right]^2}\right]+\nonumber\\
&\quad\qquad+\theta^{\mu\t}\left(\frac{g_{\t\s}}{\tilde{k}^2}-2\frac{\tilde{k}_\t\tilde{k}_\s}{\tilde{k}^4}\right)\theta^{\s\nu}\bigg\}=\\
i\Pi^{\mu\nu}_{IR}(p)&=4g^2\int\frac{d^dk}{(2\pi)^d}\left\{(d-3)\left(2\frac{k^\mu k^\nu}{k^4}-\frac{g^{\mu\nu}}{k^2}\right)+\theta^{\mu\t}\left(\frac{g_{\t\s}}{\tilde{k}^2}-2\frac{\tilde{k}_\t\tilde{k}_\s}{\tilde{k}^4}\right)\theta^{\s\nu}\right\}\sin^2\left(\frac{k\tilde{p}}{2}\right).
\end{align}
\end{appendix}

\end{document}